\documentclass[acmsmall]{acmart}

\makeatletter
\renewcommand*{\p@section}{\S}
\renewcommand*{\p@subsection}{\S}
\renewcommand*{\p@subsubsection}{\S}
\makeatother

\usepackage{etoolbox}
\AtBeginEnvironment{tabular}{\small}

\AtBeginDocument{%
  \providecommand\BibTeX{{%
    \normalfont B\kern-0.5em{\scshape i\kern-0.25em b}\kern-0.8em\TeX}}}

\setcopyright{acmcopyright}
\acmJournal{TOSN}
\acmYear{2020} \acmVolume{1} \acmNumber{1} \acmArticle{1} \acmMonth{1} \acmPrice{15.00}\acmDOI{10.1145/3409477}

\usepackage[export]{adjustbox}
\usepackage{booktabs} 
\usepackage{multirow} 
\usepackage{siunitx} 
\usepackage{tablefootnote}
\usepackage{paralist} 
\usepackage[utf8]{inputenc} 
\usepackage[T1]{fontenc} 
\usepackage[normalem]{ulem} 
\usepackage{xspace}
\usepackage{layouts} 
\usepackage{wrapfig} 
\usepackage{xfrac} 
\usepackage{enumitem} 
\usepackage{layouts} 
\usepackage{color}
\usepackage{balance}

\usepackage[caption=false, nearskip=2pt]{subfig}

\usepackage{tikz}
\usetikzlibrary{arrows}

\usepackage[ruled,vlined,linesnumbered]{algorithm2e}

\newcommand{\fakeparagraphnospace}[1]{\noindent\textbf{#1.}}
\newcommand{\fakeparagraph}[1]{\fakeparagraphnospace{#1}}

\usepackage{mathtools} 
\usepackage{amsmath}
\DeclareMathOperator*{\argmin}{arg\,min}

\newcommand{\norm}[1]{\left\lVert#1\right\rVert}

\newcommand{\vs}{vs.\ }

\newcommand{\wrt}{w.r.t.\ }
\newcommand{\ie}{i.e.,\ }
\newcommand{\eg}{e.g.,\ }

\newcommand{\decawave}{Decawave\xspace}

\newcommand{\poll}{\textsc{poll}\xspace}
\newcommand{\response}{\textsc{response}\xspace}

\newcommand{\sstwr}{SS-TWR\xspace}
\newcommand{\dstwr}{DS-TWR\xspace}
\newcommand{\deltat}{\ensuremath{\Delta t}\xspace}

\newcommand{\deltad}{\ensuremath{\Delta d}\xspace}
\newcommand{\dtx}{\ensuremath{T_\mathit{RESP}}\xspace}
\newcommand{\dtxi}{\ensuremath{T_\mathit{RESP,i}}\xspace}

\newcommand{\tprx}{\ensuremath{t_2}\xspace}
\newcommand{\trtx}{\ensuremath{t_3}\xspace}

\newcommand{\tof}{\ensuremath{\tau}\xspace}
\newcommand{\tofi}{\ensuremath{\tau_\mathit{i}}\xspace}
\newcommand{\prr}{\ensuremath{\mathit{PRR}}\xspace}
\newcommand{\aggprr}{\ensuremath{\overline{\mathit{PRR}}}\xspace}
\newcommand{\nth}[1]{#1\ensuremath{^\mathit{th}}\xspace}
\newcommand{\ieeestd}{IEEE~802.15.4\xspace}
\newcommand{\prf}{\ensuremath{\mathit{PRF}}\xspace}
\newcommand{\prfs}{\ensuremath{\mathit{PRFs}}\xspace}

\newcommand{\rtt}{\ensuremath{T_\mathit{RTT}}\xspace}
\newcommand{\rtti}{\ensuremath{T_\mathit{RTT,i}}\xspace}
\newcommand{\atx}{\ensuremath{A_\mathit{TX}}\xspace}
\newcommand{\indexi}{\ensuremath{\Tau_i}\xspace}

\newcommand{\indexfpi}{\ensuremath{\Tau_\mathit{FP,i}}\xspace}
\newcommand{\tfp}{\ensuremath{T_\mathit{FP}}\xspace}
\newcommand{\centerpos}{\textsc{center}\xspace}
\newcommand{\edgepos}{\textsc{edge}\xspace}

\newcommand{\niterations}{\ensuremath{K}\xspace}
\newcommand{\threshold}{\ensuremath{\eta}\xspace}
\newcommand{\fpindex}{\texttt{FP\_INDEX}\xspace}

\newcommand{\toa}{ToA\xspace}
\newcommand{\tdoa}{TDoA\xspace}
\newcommand{\ssub}{S{\footnotesize \&}S\xspace}

\newcommand{\crng}{concurrent ranging\xspace}
\newcommand{\resp}[1]{\ensuremath{R_{#1}}\xspace}
\newcommand{\rmarker}{\texttt{RMARKER}\xspace}
\newcommand{\stdnoise}{\ensuremath{\sigma_n}\xspace}
\newcommand{\tdelay}{\ensuremath{T_\mathit{ID}}\xspace}
\newcommand{\trimstep}{\ensuremath{\mathcal{S}}\xspace}
\newcommand{\Tau}{\mathcal{T}}
\newcommand{\detuning}{\ensuremath{T_\mathit{det}}\xspace}

\begin{document}

\title{Ultra-wideband Concurrent Ranging}

\author{Pablo Corbal\'an}
\email{p.corbalanpelegrin@unitn.it}
\affiliation{
  \institution{University of Trento}
  \country{Italy}
}
\author{Gian Pietro Picco}
\email{gianpietro.picco@unitn.it}
\affiliation{
  \institution{University of Trento}
  \country{Italy}
}

\renewcommand{\shortauthors}{Corbal\'an and Picco}

\begin{abstract}
  We propose a novel \emph{concurrent ranging} technique for distance
  estimation with ultra-wideband (UWB) radios.  Conventional schemes
  assume that the necessary packet exchanges occur in isolation, to
  avoid collisions. Concurrent ranging relies on the
  \emph{overlapping} of replies from nearby responders to the same
  ranging request issued by an initiator node. As UWB transmissions
  rely on short pulses, the individual times of arrival can be
  discriminated by examining the channel impulse response (CIR) of the
  initiator transceiver. By ranging against $N$ responders with a
  \emph{single}, concurrent exchange, our technique drastically abates
  network overhead, enabling higher ranging frequency with lower
  latency and energy consumption \wrt conventional schemes.

  Concurrent ranging can be implemented with a strawman approach
  requiring minimal changes to standard schemes. Nevertheless, we
  empirically show that this limits the attainable accuracy,
  reliability, and therefore applicability. We identify the main
  challenges in realizing concurrent ranging without dedicated
  hardware and tackle them by contributing several techniques, used in
  synergy in our prototype based on the popular DW1000
  transceiver. Our evaluation, with static targets and a mobile robot,
  confirms that concurrent ranging reliably achieves decimeter-level
  distance and position accuracy, comparable to conventional schemes
  but at a fraction of the network and energy cost.
\end{abstract}

\ccsdesc[500]{Networks~Location based services}

\keywords{Ultra-wideband, Concurrent Transmissions, Ranging, Localization}


\maketitle

\section{Introduction}
\label{sec:intro}

A new generation of localization systems is rapidly gaining interest,
fueled by countless applications~\cite{benini2013imu,follow-me-drone,
museum-tracking, mattaboni1987autonomous, guo2016ultra, irobot-lawnmower,
  fontana2003commercialization} for which global navigation satellite
systems do not provide sufficient reliability, accuracy, or update
rate. These so-called real-time location systems (RTLS) rely on
several technologies, including optical~\cite{optitrack, vicon},
ultrasonic~\cite{cricket, alps, ultrasonic-tdoa},
inertial~\cite{benini2013imu}, and radio frequency (RF). Among these,
RF is predominant, largely driven by the opportunity of exploiting
ubiquitous wireless communication technologies like WiFi and Bluetooth
also towards localization. Localization systems based on these radios
enjoy, in principle, wide applicability; however, they typically
achieve meter-level accuracy, enough for several use cases but
insufficient for many others.

Nevertheless, another breed of RF-based localization recently 
re-emerged from a decade-long oblivion: ultra-wideband (UWB). 
The recent availability of tiny, low-cost, and low-power 
UWB transceivers has renewed interest in this technology, 
whose peculiarity is to enable accurate distance
estimation (\emph{ranging}) along with high-rate communication. These
characteristics are rapidly placing UWB in a dominant position in the
RTLS arena, and defining it as a key enabler for several Internet of
Things (IoT) and consumer scenarios. UWB is currently not as
widespread as WiFi or BLE, but the fact that the latest 
Apple iPhone~11 is equipped with a UWB transceiver is a witness 
that the trend may change dramatically in the near future.

The \decawave DW1000 transceiver~\cite{dw1000-datasheet} has been at
the forefront of this technological advancement, as it provides
centimeter-level ranging accuracy with a tiny form factor and a power
consumption an order of magnitude lower than its bulky UWB
predecessors. On the other hand, this consumption is still an order of
magnitude higher than other IoT low-power wireless radios; 
further, its impact is exacerbated when ranging---the key 
asset of UWB---is exploited, due to the long packet exchanges required.

\begin{figure}[!t]
\centering
\subfloat[Single-sided two-way ranging (\sstwr).\label{fig:two-way-ranging}]{
  \includegraphics[width=.49\textwidth, valign=t]{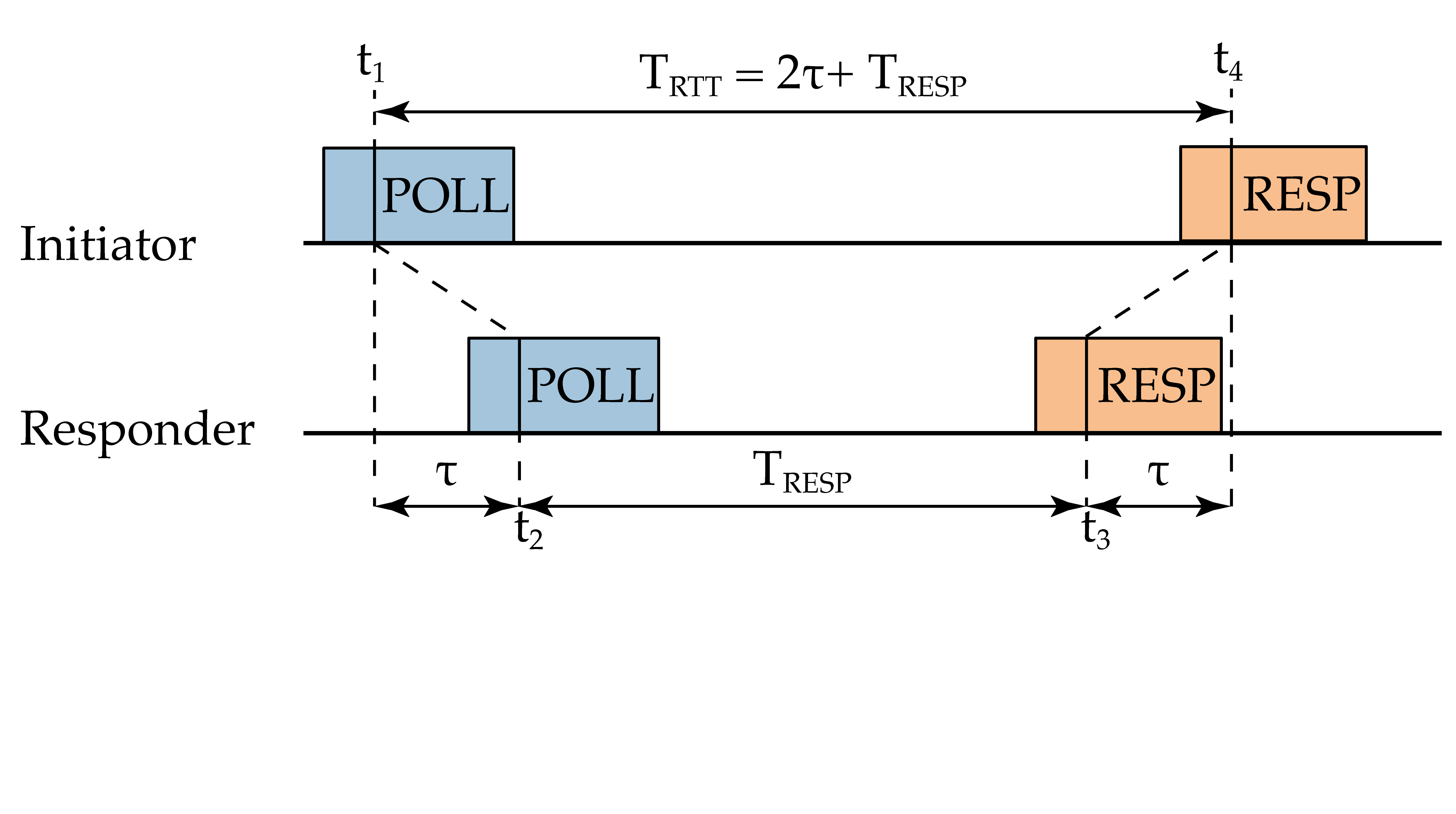}}
\hfill
\subfloat[Concurrent ranging.\label{fig:crng}]{
  \includegraphics[width=.49\textwidth, valign=t]{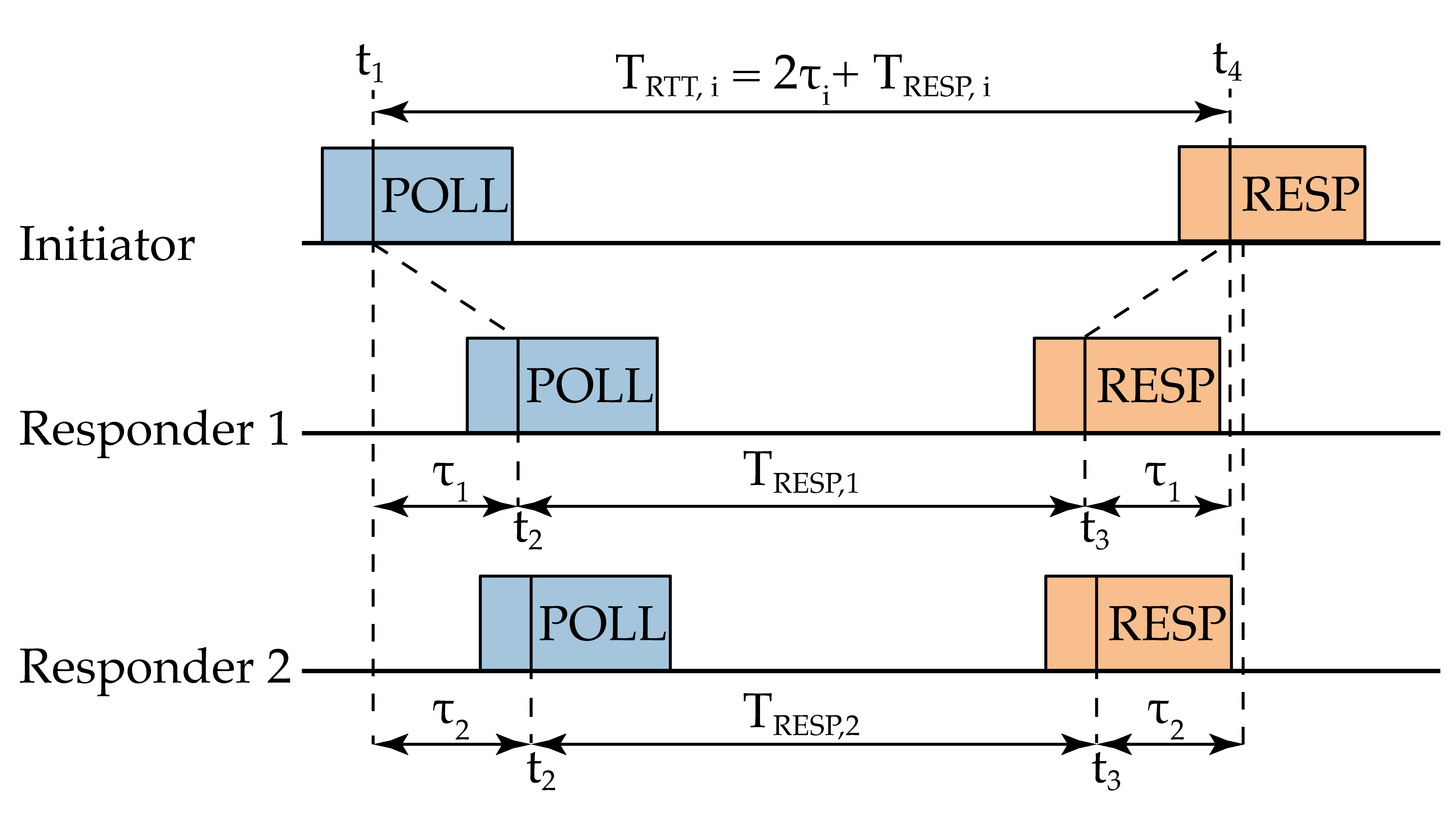}}
\caption{In \sstwr, the initiator transmits a unicast
  \poll to which a single responder replies with a \response. In 
  concurrent ranging, the initiator transmits a \emph{broadcast} 
  \poll to which responders in range reply concurrently.}
\label{fig:sstwr-crng-cmp}
\end{figure}

\fakeparagraph{UWB Two-way Ranging (TWR)} 
Figure~\ref{fig:two-way-ranging} illustrates 
single-sided two-way ranging (SS-TWR), the simplest 
scheme, part of the IEEE~802.15.4-2011
standard~\cite{std154} and further illustrated in~\ref{sec:background}. 
The \emph{initiator}\footnote{The IEEE
  standard uses \emph{originator} instead of \emph{initiator}; we
  follow the terminology used by the \decawave documentation.} requests
a ranging measurement via a \poll packet; the responder, after a known
delay \dtx, replies with a \response packet containing the timestamps
marking the receipt of \poll and the sending of \response. This
information, along with the dual timestamps marking the sending of
\poll and the receipt of \response measured locally at the initiator, 
enable the latter to accurately compute the 
time~of~flight $\tau$ 
and estimate the distance from the responder as $d=\tof \times c$,
where $c$ is the speed of light in air. 

Two-way ranging, as the name suggests, involves a \emph{pairwise}
exchange between the initiator and \emph{every} responder. In other
words, if the initiator must estimate its distance \wrt $N$ nodes,
$2\times N$ packets are required.  The situation is even worse with
other schemes that improve accuracy by acquiring more timestamps via
additional packet transmissions, \eg up to $4\times N$ in 
popular double-sided two-way ranging 
(\dstwr) schemes~\cite{dstwr, dw-dstwr-patent, dw-dstwr}. 

\fakeparagraph{UWB Concurrent Ranging} 
We propose a novel approach to ranging in which, 
instead of \emph{separating} the pairwise exchanges necessary 
to ranging, these are \emph{overlapping} in time (Figure~\ref{fig:crng}).
Its mechanics are extremely simple: when the single
(broadcast) \poll sent by the initiator is received, 
each responder sends back its \response as if it were alone,
effectively yielding concurrent replies to the initiator. 
This \emph{concurrent ranging} technique enables the initiator 
to \emph{range with $N$ nodes at once by using only 2~packets}, 
\ie as if it were ranging against a single responder. 
This significantly reduces latency and energy consumption, 
increasing scalability and battery lifetime,
but causes the concurrent signals from different
responders to ``fuse'' in the communication channel, potentially
yielding a collision at the initiator.

This is precisely where the peculiarities of UWB communications
come into play. UWB transmissions rely on very
short ($\leq$2~ns) pulses, enabling very precise timestamping of 
incoming radio signals. This is what makes UWB intrinsically more
amenable to accurate ranging than narrowband, whose reliance on
carrier waves that are more ``spread in time'' induces physical bounds
on the precision that can be attained in establishing a time reference
for an incoming signal. 
Moreover, it is what enables our novel idea of
concurrent ranging. In narrowband, the fact that concurrent signals
are spread over time makes them very difficult to tell apart once
fused into a single signal. In practice, this is possible only if
detailed channel state information is available---usually not the case
on narrowband low-power radios, \eg the popular CC2420~\cite{cc2420} 
and its recent descendants.  In contrast, the reliance of UWB 
on short pulses makes concurrent signals less likely to collide and combine
therefore enabling, under certain conditions discussed later, 
their identification if channel impulse response (CIR) information is available. 
Interestingly, the DW1000 
\begin{inparaenum}[\itshape i)]
\item bases its own operation precisely on the processing of the CIR, and 
\item makes the CIR available also to the application layer (\ref{sec:background}).
\end{inparaenum}

\fakeparagraph{Goals and Contributions}
As discussed in~\ref{sec:crng}, a strawman implementation of \crng is
very simple. Therefore, using our prototype deployed in a small-scale
setup, we begin by investigating the \emph{feasibility} of \crng
(\ref{sec:questions}), given the inevitable degradation in accuracy
\wrt isolated ranging caused by the interference among the signals of
responders, in turn determined by their relative placement. Our
results, originally published in~\cite{crng},
offer empirical evidence that it is indeed possible to derive accurate
ranging information from UWB signals overlapping in time.

On the other hand, these results also point out the significant
\emph{challenges} that must be overcome to transform \crng from an
enticing opportunity to a practical system. Solving these
  challenges is the specific goal of this paper \wrt the original
  one~\cite{crng} where, for the first time in the literature, we have
  introduced the concept and shown the feasibility of concurrent
  ranging.

Among these challenges, a key one is the \emph{limited precision of scheduling
  transmissions} in commercial UWB transceivers.  For instance, the
popular \decawave DW1000 we use in this work can timestamp packet receptions (RX) 
with a precision of $\approx$15~ps, but can schedule
transmissions (TX) with a precision of only $\approx$8~ns. This is
not an issue in conventional ranging schemes like \sstwr; as mentioned
above, the responder embeds the necessary timestamps in the \response
payload, allowing the initiator to correct for the limited TX
granularity. However, in \crng only one \response is decoded, 
if any; the timing information of the others must be
derived solely from the appearance of their corresponding signal 
paths in the CIR. This process is greatly affected by the TX uncertainty, 
which significantly reduces accuracy and consequently 
hampers the practical adoption of \crng.

In this paper, 
we tackle and solve this key challenge with a mechanism that
significantly improves the TX scheduling precision via a \emph{local}
compensation (\ref{sec:reloaded}).  Indeed, both the precise and
imprecise information about TX scheduling are available at the
responder; the problem arises because the radio discards the less
significant 9~bits of the precise 40-bit timestamp. Therefore, the
responder can correct for the \emph{known} TX timing error when
preparing its \response. We achieve this by fine-tuning the frequency
of the crystal oscillator entirely in firmware and locally to the
responder, \ie without additional hardware or external out-of-band
infrastructure. Purposely, the technique also compensates for the
oscillator frequency offset between initiator and responders,
significantly reducing the impact of clock drift, the main cause of
ranging error in \sstwr.

Nevertheless, precisely scheduling transmissions 
is not the only challenge of \crng. A full-fledged, 
practically usable system also requires tackling
\begin{inparaenum}[\itshape i)]
\item the reliable identification of the concurrent responders, and
\item the precise estimation of the time of arrival (\toa) of their signals;
\end{inparaenum}
both are complicated by the intrinsic mutual interference of
concurrent transmissions. In this paper, we build upon techniques developed
by us~\cite{chorus} and other groups~\cite{crng-graz,snaploc} since we
first proposed concurrent ranging in~\cite{crng}. Nevertheless, we
\emph{adapt and improve} these techniques (\ref{sec:reloaded}) to
accommodate the specifics of concurrent ranging in general and 
the TX scheduling compensation technique in particular. 
Interestingly, our novel design significantly increases 
not only the accuracy but also the \emph{reliability} of 
\crng \wrt our original strawman design in~\cite{crng}. 
The latter relied heavily 
\begin{inparaenum}[\itshape i)]
\item on the successful RX of at least one \response, 
  containing the necessary timestamps for
  accurate time-of-flight calculation, and 
\item on the \toa estimation of this \response performed by the DW1000, 
  used to determine the difference in the signal \toa 
  (and therefore distance) to the other responders. 
\end{inparaenum}
However, the fusion of concurrent signals may cause 
the decoding of the \response to be matched to the
wrong responder or fail altogether, yielding grossly incorrect
estimates or none at all, respectively. Thanks to the ability to
precisely schedule the TX of \response packets, we
\begin{inparaenum}[\itshape i)]
\item remove the need to
  decode at least one of them, and
\item enable distance estimation \emph{solely} based on the CIR. 
\end{inparaenum}
We can actually \emph{remove the payload entirely} from \response packets,
further reducing latency and energy consumption.

We evaluate \crng extensively (\ref{sec:crng-tosn:eval}). We first
show via dedicated experiments that our prototype can schedule TX with
$<1$~ns error. We then analyze the \emph{raw} positioning information
obtained by \crng, to assess its quality without the help of
additional filtering techniques~\cite{ukf-julier, ukf} that, as
shown in~\cite{atlas-tdoa, guo2016ultra, 7374232, ethz-one-way}, would
nonetheless improve performance. Our experiments in two environments,
both with static positions and mobile trajectories, confirm that the
near-perfect TX scheduling precision we achieve, along with our dedicated
techniques to accurately extract distance information from the CIR,
enable reliable decimeter-level ranging and positioning
accuracy---same as conventional schemes for UWB but at a fraction of
the network and energy cost.

These results, embodied in our prototype implementation, confirm that
UWB concurrent ranging is a concrete option, immediately applicable to
real-world applications where it strikes new trade-offs \wrt accuracy,
latency, energy, and scalability, offering a valid (and often more
competitive) alternative to established conventional methods,
as discussed in~\ref{sec:discussion}.

Finally, in~\ref{sec:relwork} we place concurrent ranging in the 
context of related work, before ending in~\ref{sec:crng-tosn:conclusions}
with brief concluding remarks.

\section{Background}
\label{sec:background}

We concisely summarize the salient features of UWB radios in general
(\ref{sec:uwb}) and how they are made available by the popular DW1000
transceiver we use in this work (\ref{sec:dw1000}). Moreover, we
illustrate the \sstwr technique we build upon,
and show how it is used to perform localization (\ref{sec:toa}).

\subsection{Ultra-wideband in the \ieeestd PHY Layer}
\label{sec:uwb}

UWB communications have been originally used for military applications due to
their very large bandwidth and interference resilience to mainstream
narrowband radios. In 2002, the FCC approved the unlicensed use of UWB under
strict power spectral masks, boosting a new wave of research from industry and
academia. Nonetheless, this research mainly focused on high data rate
communications, and remained largely based on theory and simulation, as most
UWB radios available then were bulky, energy-hungry, and expensive, hindering
the widespread adoption of UWB. In 2007, the {\ieeestd}a standard amendment
included a UWB PHY layer based on impulse radio (IR-UWB)~\cite{impulse-radio}, 
aimed at providing accurate ranging with low-power consumption. 
A few years ago, \decawave released a standard-compliant IR-UWB radio, 
the DW1000, saving UWB from a decade-long oblivion, and taking 
by storm the field of real-time location systems (RTLS).

\begin{figure}[!t]
\begin{minipage}[t]{0.48\linewidth}
\centering
\includegraphics{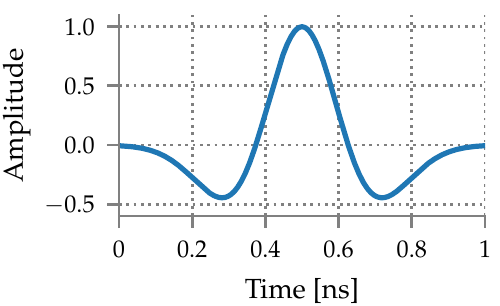}
\caption{UWB pulse.}
\label{fig:uwb-pulse}
\end{minipage}
\begin{minipage}[t]{0.51\linewidth}
\centering
\includegraphics{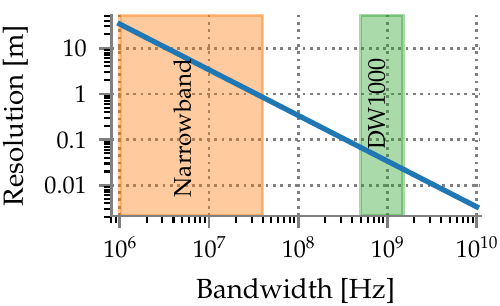}
\caption{Distance resolution \vs bandwidth.}
\label{fig:dest-bw}
\end{minipage}
\vspace{-2mm}
\end{figure}

\fakeparagraph{Impulse Radio} 
According to the FCC, UWB signals are
characterized by a bandwidth $\geq 500$~MHz or a fractional bandwidth
$\geq 20\%$ during transmission. To achieve such a large bandwidth,
modern UWB systems are based on IR-UWB, using pulses
(Figure~\ref{fig:uwb-pulse}) very narrow in time ($\leq 2$~ns). 
This reduces the power spectral density, the interference
produced to other wireless technologies, and the impact of multipath
components (MPC). Further, it enhances the ability of UWB signals to
propagate through obstacles and walls~\cite{uwb-idea} and simplifies
transceiver design.  The large bandwidth also provides excellent time
resolution (Figure~\ref{fig:dest-bw}), enabling UWB receivers to
precisely estimate the time of arrival (\toa) of a signal and
distinguish the direct path from MPC.  Time-hopping
codes~\cite{ir-uwb-maccess} enable multiple access to the
medium. Overall, these features make \mbox{IR-UWB} ideal for low-power
ranging and localization as well as communication.

\fakeparagraph{\ieeestd UWB PHY Layer} The \ieeestd-2011
standard~\cite{std154} specifies a PHY layer based on IR-UWB. 
The highest frequency at which a compliant device shall emit
pulses is 499.2~MHz (fundamental frequency), yielding a
standard chip duration of $\approx2$~ns. A UWB frame is composed of
\begin{inparaenum}[\itshape i)]
\item a synchronization header (SHR) and 
\item a data portion. 
\end{inparaenum}
The SHR is encoded in single pulses and includes a preamble for
synchronization and the start frame delimiter (SFD), which delimits
the end of the SHR and the beginning of the data portion.
Instead, the data portion exploits a combination of burst 
position modulation (BPM) and binary phase-shift keying (BPSK), and 
includes a physical header (PHR) and the data payload. 
The duration of the preamble is configurable and depends on 
the number of repetitions of a predefined symbol, whose structure 
is determined by the preamble code. Preamble codes also define the
pseudo-random sequence used for time-hopping in the transmission of
the data part. The standard defines preamble codes of $31$ and $127$
elements, which are then interleaved with zeros according to a
spreading factor. This yields a (mean) \emph{pulse repetition
  frequency} (\prf) of $16$~MHz or $64$~MHz.  Preamble codes and \prfs
can be exploited to configure non-interfering links within the same RF
channel~\cite{uwb-ctx-fire}.

\subsection{\decawave DW1000}
\label{sec:dw1000}

The \decawave DW1000~\cite{dw1000-datasheet} is a commercially
available low-power low-cost UWB transceiver compliant with \ieeestd,
for which it supports frequency channels 1--4 in the low band and 5, 7
in the high band, and data rates of $110$~kbps, $850$~kbps, and
$6.8$~Mbps. Channels 4 and~7 have a larger $900$~MHz bandwidth, while
the others are limited to $499.2$~MHz.

\fakeparagraph{Channel Impulse Response (CIR)} The perfect periodic
autocorrelation of the preamble code sequence enables coherent
receivers to determine the CIR~\cite{dw1000-manual-v218}, which provides
information about the multipath propagation characteristics of the
wireless channel between a transmitter and a receiver.  The CIR allows
UWB radios to distinguish the signal leading edge,
commonly called\footnote{Hereafter, 
  we use the terms first path and direct path interchangeably.} 
\emph{direct} or \emph{first} path, 
from MPC and accurately estimate the \toa of the signal. 
In this paper, we exploit the information available 
in the CIR to perform these operations on \emph{several signals 
transmitted concurrently}.

The DW1000 measures the CIR upon preamble reception with a sampling
period \mbox{$T_s = 1.0016$~ns}. The CIR is stored in a large internal
buffer of 4096B accessible by the firmware developer.  The time span
of the CIR is the duration of a preamble symbol: 992~samples for a
16~MHz \prf or 1016 for a 64~MHz \prf. Each sample is a complex number
$a_k + jb_k$ whose real and imaginary parts are 16-bit signed
integers.  The amplitude $A_k$ and phase $\theta_k$ at each time delay
$t_k$ is $A_k = \sqrt{\smash[b]{a_k^2 + b_k^2}}$ and
$\theta_k = \arctan{\frac{b_k}{a_k}}$. The DW1000 measures the CIR
even when RX errors occur, therefore offering signal timing
information even when a packet (\eg a \response) cannot be
successfully decoded.

\fakeparagraph{TX/RX Timestamps} The TX and RX timestamps enabling ranging
are measured in a packet at the ranging marker (\rmarker)~\cite{dw1000-manual-v218}, 
which marks the first pulse of the PHR after the SFD (\ref{sec:uwb}). 
These timestamps are measured with a very high time resolution 
in radio units of $\approx\SI{15.65}{\pico\second}$. 
The DW1000 first makes a coarse RX timestamp estimation, 
then adjusts it based on
\begin{inparaenum}[\itshape i)]
\item the RX antenna delay, and 
\item the first path in the CIR estimated by a proprietary
  internal leading edge detection (LDE) algorithm. 
\end{inparaenum}
The CIR index that LDE determines to be the first path
  (\fpindex) is stored together with the RX timestamp in the
  \texttt{RX\_TIME} register. LDE detects the first path as the
first sampled amplitude that goes above a dynamic threshold based on
\begin{inparaenum}[\itshape i)]
\item the noise standard deviation \stdnoise and 
\item the noise peak value.
\end{inparaenum}
Similar to the CIR, the RX signal timestamp is measured despite RX errors,
unless there is a rare PHR error~\cite[p. 97]{dw1000-manual-v218}. 

\fakeparagraph{Delayed Transmissions} The DW1000 offers the capability to
schedule transmissions at a specified time in the
future~\cite[p. 20]{dw1000-manual-v218}, corresponding to the
\rmarker. To this end, the DW1000 internally computes the time at
which to begin the preamble transmission, considering also the TX
antenna delay~\cite{dw1000-antenna-delay}. This makes the TX
timestamp predictable, which is key for ranging.

\begin{table}[!tb]
  \caption{Current consumption comparison of DW1000 \vs TI CC2650 BLE
    SoC~\cite{cc2650-datasheet} and Intel 5300 WiFi
    card~\cite{wifi-power}.  Note that the CC2650 includes a 32-bit
    ARM Cortex-M3 processor and the Intel~5300 can support multiple
    antennas; further, consumption depends on radio configuration.}
  \label{tab:current-consumption}
  \begin{tabular}{l c c c}
    \toprule
     & \textbf{DW1000} & \textbf{TI CC2650~\cite{cc2650-datasheet}}& \textbf{Intel 5300~\cite{wifi-power}}\\
    \textbf{State} & 802.15.4a & BLE~4.2 \& 802.15.4 & 802.11~a/b/g/n\\
    \midrule
    Deep Sleep & 50~\si{\nano\ampere} & 100--150~\si{\nano\ampere} & N/A\\
    Sleep & 1~\si{\micro\ampere} & 1~\si{\micro\ampere} & 30.3 ~mA\\
    Idle & 12--18~mA & 550~\si{\micro\ampere} & 248~mA\\
    TX & 35--85~mA & 6.1--9.1~mA & 387--636~mA\\
	RX & 57--126~mA & 5.9--6.1~mA & 248--484~mA\\
  \bottomrule
  \end{tabular}
\end{table}

\fakeparagraph{Power Consumption} 
An important aspect of the DW1000 is its
low-power consumption \wrt previous UWB
transceivers (\eg~\cite{timedomain-pulson400}).
Table~\ref{tab:current-consumption} compares the current consumption of
the DW1000 against other commonly-used technologies (BLE and WiFi) for
localization. The DW1000 consumes significantly less than the Intel
5300~\cite{wifi-power}, which provides channel state information
(CSI).  However, it consumes much more than low-power
widespread technologies such as BLE or
\ieeestd~narrowband~\cite{cc2650-datasheet}.  Hence, to ensure a long
battery lifetime of UWB devices it is essential to reduce the
radio activity, while retaining the accuracy and update rate of
ranging and localization required by applications.

\subsection{Time-of-Arrival (ToA) Ranging and Localization}
\label{sec:toa}

In \toa-based methods, distance is estimated by precisely measuring
RX and TX timestamps of packets exchanged between nodes. In this section, 
we describe the popular \sstwr ranging technique (\ref{sec:soa-sstwr}) 
we extend and build upon in this paper, 
and show how distance estimates from known positions can be used to
determine the position of a target (\ref{sec:soa:toa-loc}).

\subsubsection{Single-sided Two-way Ranging (\sstwr)}
\label{sec:soa-sstwr}

In \sstwr, part of the \ieeestd standard~\cite{std154}, the initiator
transmits a unicast \poll packet to the responder, storing the TX
timestamp $t_1$ (Figure~\ref{fig:two-way-ranging}). The responder
replies back with a \response packet after a given response delay
\dtx. Based on the corresponding RX timestamp $t_4$, the initiator can
compute the round trip time $\rtt = t_4 - t_1 = 2\tau +
\dtx$. However, to cope with the limited TX scheduling precision of
commercial UWB radios, the \response payload includes the RX timestamp
$t_2$ of the \poll and the TX timestamp $t_3$ of the \response,
allowing the initiator to precisely measure the actual response delay
$\dtx = t_3 - t_2$.  The time of flight $\tau$ can be then computed as
\begin{equation*}\label{eq:sstwr-tof}
\tau = \frac{\rtt - \dtx}{2} = \frac{(t_4 - t_1) - (t_3 - t_2)}{2} 
\end{equation*}
and the distance between the two nodes estimated as
$d = \tau \times c$, where $c$ is the speed of light in air.

\sstwr is simple, yet provides accurate distance estimation for many
applications. The main source of error is the clock drift between
initiator and responder, each running an internal oscillator with an
offset \wrt the expected nominal frequency~\cite{dw-errors}, 
causing the  actual time of flight measured by the initiator to be
\begin{equation*}
\hat{\tau} = \frac{\rtt(1+e_I) - \dtx(1+e_R)}{2}
\end{equation*}
where $e_I$ and $e_R$ are the crystal offsets of initiator and
responder, respectively. After some derivations, and by observing that
$\dtx \gg 2\tau$, we can approximate the error
to~\cite{dstwr,dw-errors}
\begin{equation*}\label{eq:sstwr-drift}
\hat{\tau} - \tau \approx \frac{1}{2} \dtx(e_I - e_R)
\end{equation*}

Therefore, to reduce the ranging error of \sstwr one should
\begin{inparaenum}[\itshape i)]
\item compensate for the drift, and
\item minimize \dtx, as the error grows linearly with it. 
\end{inparaenum}

\subsubsection{Position Estimation}
\label{sec:soa:toa-loc}

The estimated distance $\hat{d_i}$ to each of the $N$ responders can be
used to determine the unknown initiator position $\mathbf{p}$, provided the
responder positions are known.  In two-dimensional space, the
Euclidean distance $d_i$ to responder \resp{i} is defined by
\begin{equation}\label{eq:soa:dist-norm}
d_i = \norm{\mathbf{p} - \mathbf{p_i}} = \sqrt{(x - x_i)^2 + (y - y_i)^2}
\end{equation}
where $\mathbf{p_i} = [x_i, y_i]$ is the position of \resp{i}, 
$i \in [1, N]$. The geometric representation of
Eq.~\eqref{eq:soa:dist-norm} is a circle (a sphere in~3D) with radius
$d_i$ and center in $\mathbf{p_i}$.  In the absence of noise, the
intersection of $N \geq 3$ circles yields the unique initiator
position $\mathbf{p}$. In practice, however, each distance estimate
$\hat{d_i} = d_i + n_i$ suffers from an additive zero-mean measurement
noise $n_i$. An estimate $\mathbf{\hat p}$ of the unknown initiator
position can be determined (in 2D) by minimizing the non-linear
least-squares (NLLS) problem
\begin{equation*}\label{eq:toa-solver-dist}
\mathbf{\hat p} = \argmin_{\mathbf{p}} 
	\sum_{i = 1}^{N}\left(\hat d_i  - \sqrt{(x - x_i)^2 + (y - y_i)^2}\right)^2
\end{equation*}
In this paper, we solve the NLLS problem with state-of-the-art
methods, as our contribution is focused on ranging and not on the
computation of the position.  Specifically, we employ an iterative
local search via a trust region reflective
algorithm~\cite{branch1999subspace}.  This requires an initial
position estimate $\mathbf{p_0}$ that we set as the solution of a
linear least squares estimator that linearizes the system of equations
by applying the difference between any two of 
them~\cite{source-loc-alg, toa-lls}.

\section{Concurrent Ranging}
\label{sec:crng}

Ranging against $N$ responders (\eg anchors) with \sstwr requires $N$
independent pairwise exchanges---essentially, $N$ instances of
Figure~\ref{fig:two-way-ranging}, one after the other. In contrast,
the notion of concurrent ranging we propose obtains the same
information within a \emph{single} exchange, as shown in
Figure~\ref{fig:crng} with only two responders. The
technique is conceptually very simple, and consists of changing the
basic \sstwr scheme (\ref{sec:soa-sstwr}) by:
\begin{compactenum}
\item replacing the $N$ unicast \poll packets necessary to solicit
  ranging from the $N$ responders with a \emph{single} broadcast \poll, and
\item having all responders reply to the \poll after the \emph{same}
  time interval \dtx from its (timestamped) receipt.
\end{compactenum}

\begin{figure}[!t]
\centering
\subfloat[Narrowband.\label{fig:concurrent-narrowband}]{
  \includegraphics{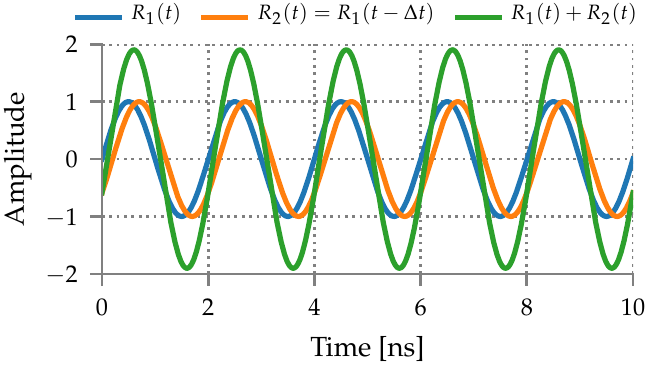}}
\hfill
\subfloat[UWB.\label{fig:concurrent-uwb}]{
  \includegraphics{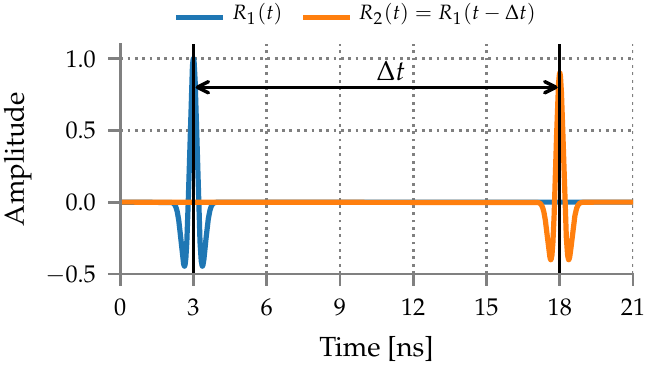}}
\caption{Concurrent ranging, idealized, with 
  narrowband (\ref{fig:concurrent-narrowband}) and 
  UWB (\ref{fig:concurrent-uwb}) radios.
  With narrowband it is infeasible to recover the timing
  information of the signals from the individual responders. With UWB, instead,
  the different distance from the initiator to responders $R_1$ and $R_2$ 
  produces a time shift \deltat between their signals. 
  By measuring \deltat, we can determine the distance 
  difference ${\deltad = \mid d_1 - d_2 \mid}$ 
  between responders.
}
\end{figure}

This simple idea is impractical (if not infeasible) in narrowband
radios. As illustrated in the idealized view of
Figure~\ref{fig:concurrent-narrowband}, the signals from responders
$R_1$ and $R_2$ interfere with each other, yielding a combined 
signal where the time information necessary to estimate distance is
essentially lost. In contrast, Figure~\ref{fig:concurrent-uwb} shows
why this is not the case in UWB; the time displacement \deltat of 
pulses from responders, caused by their different distances from
the initiator, is still clearly visible in the resulting signal. This
is due to the fact that UWB pulses are extremely short \wrt narrowband
waves, and therefore unlikely to interfere---although in practice
things are not so simple, as discussed in~\ref{sec:questions}.

\fakeparagraph{A Strawman Implementation} Concurrent ranging can be
implemented very easily, a direct consequence of the simplicity of the
concept. If a \sstwr implementation is already available, it suffices to
replace the unicast \poll with a broadcast one. The computation of the
actual ranging estimate requires processing the available CIR signal. 
The time shift \deltat can be measured as the difference
between the first path from the closest responder \resp{1} in the CIR,
automatically obtained from the DW1000 and used to compute the
accurate distance estimate $d_1$, and the first path 
from \resp{2}, that must be instead determined in a custom way, as discussed
later (\ref{sec:cir-enough}). Indeed, the first path from \resp{2}, key to the
operation of concurrent ranging, is treated as MPC or noise 
by the DW1000 but remains visible in the CIR, enabling
computation of its time of arrival (\toa).  Once \deltat is
determined, the spatial displacement $\deltad=c\deltat$ can be
computed, along with the distance $d_2 = d_1 + \deltad$ of \resp{2}; 
a similar process must be repeated in the case of $N$ responders.

As for the value of the response delay, 
crucial to the accuracy of \sstwr (\ref{sec:soa-sstwr}), our implementation uses
$\dtx=330$~\si{\micro\second}. We verified experimentally that this
provides a good trade-off; lower values do not leave enough time to
correctly prepare the radio for the \response transmission,
and larger ones negatively affect ranging due to clock drift. 

In \crng, as in \sstwr, the \dtx value also enables the responder 
to determine the time \mbox{$\trtx = \tprx + \dtx$} at which the \response 
must be sent (Figure~\ref{fig:sstwr-crng-cmp}). 
The timestamp \tprx associated to the RX of \poll is estimated
by the DW1000 at the \rmarker with the extremely high precision of
15~ps (\ref{sec:uwb}).  Unfortunately, the same precision is not
available when scheduling the delayed TX (\ref{sec:dw1000}) of the
corresponding \response at time \trtx. Due to the significantly
coarser granularity of TX scheduling in the DW1000, the TX of a
\response expected at a time $\trtx$ actually 
occurs at $t + \epsilon$, 
with $\epsilon \in [-8,0)$~ns~\cite{dw1000-datasheet}. 
This is not a problem in \sstwr,
as the timestamps \tprx and \trtx are embedded in the \response and
decoded by the initiator. Instead, in \crng the additional \response
packets are not decoded, and this technique cannot be used. Therefore,
the uncertainty of TX scheduling, which at first may appear a
negligible hardware detail, has significant repercussions on the
practical exploitation of our technique, as discussed next.

\section{Feasibility and Challenges: Empirical Observations}
\label{sec:questions}
Although the idea of concurrent ranging is extremely simple and can be
implemented straightforwardly on the DW1000, several questions must be
answered to ascertain its practical feasibility.  We discuss them
next, providing answers based on empirical observations.

\subsection{Experimental Setup}
\label{sec:ewsn:setup}

All our experiments employ the \decawave EVB1000 development
platform~\cite{evb1000}, equipped with the DW1000 transceiver, an
STM32F105 ARM Cortex-M3 MCU, and a PCB antenna.

\fakeparagraph{UWB Radio Configuration} We use a preamble length of
128~symbols and a data rate of 6.8~Mbps. Further, we use channel~4,
whose wider bandwidth provides better resolution in determining 
the timing of the direct path and therefore better ranging estimates.

\fakeparagraph{Firmware} We program the behavior of initiator and responder
nodes directly atop \decawave libraries, without any OS layer, by adapting
towards our goals the demo code provided by \decawave. Specifically, we provide
support to log, via the USB interface, 
\begin{inparaenum}[\itshape i)]
\item the packets transmitted and received,
\item the ranging measurements, and
\item the CIR measured upon packet reception.
\end{inparaenum}

\fakeparagraph{Environment} All our experiments are carried out in a
university building, in a long corridor whose width
is 2.37~m. This is arguably a challenging environment due to the
presence of strong multipath, but also very realistic to test the 
feasibility of \crng, given that one of the main applications of 
UWB is for localization in indoor environments.

\fakeparagraph{Network Configuration} 
In all experiments, one initiator node and one or more responders
are arranged in a line, placed exactly in the middle of the
aforementioned corridor. This one-dimensional configuration 
allows us to clearly and intuitively relate the temporal
displacements of the received signals to the spatial displacement of
their source nodes. For instance, Figure~\ref{fig:capture-exp-deployment}
shows the network used in~\ref{sec:obs:prr}; we change
the arrangement and number of nodes depending
on the question under investigation.

\subsection{Is Communication Even Possible?}
\label{sec:obs:prr}

Up to this point, we have implicitly assumed that the UWB transceiver
is able to successfully decode one of the concurrent TX with high
probability, similarly to what happens in narrowband and exploited,
\eg by Glossy~\cite{glossy} and other protocols~\cite{lwb, chaos, crystal}. 
However, this may not be the case, given the different radio PHY and 
the different degree of synchronization (ns \vs $\mu$s) involved.

\begin{figure}[!tb]
\centering
\begin{tikzpicture}[xscale=0.6]
\tikzstyle{device} = [circle, thick, draw=black, fill=gray!30, minimum size=8mm]
\node[device] (a) at (0, 0) {$R_1$};
\node[device] (b) at (4, 0) {$I$};
\node[device] (c) at (10, 0) {$R_2$};
\draw[<->, thick] (a) -- (b) node[pos=.5,above]{$d_1$};
\draw[<->, thick] (b) -- (c) node[pos=.5,above]{$d_2 = D - d_1$};
\draw[<->, thick] (0, -.5) -- (10, -.5) node[pos=.5,below]{$D = 12$~m};
\end{tikzpicture}
\caption{Experimental setup to investigate the reliability and accuracy of
  concurrent ranging (\ref{sec:obs:prr}--\ref{sec:obs:accuracy}). $I$ is the
initiator, $R_1$ and $R_2$ are the responders.}
\label{fig:capture-exp-deployment}
\end{figure}
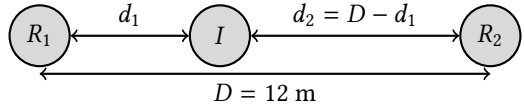

Our first goal is therefore to verify this hypothesis. We run a series
of experiments with three nodes, one initiator $I$ and two concurrent
responders $R_1$ and $R_2$, placed along a line
(Figure~\ref{fig:capture-exp-deployment}). The initiator is placed
between responders at a distance $d_1$ from $R_1$ and $d_2 = D - d_1$
from $R_2$, where $D = 12$~m is the fixed distance between the
responders. We vary $d_1$ between 0.4~m and 11.6~m in steps of
0.4~m. By changing the distance between initiator and 
responders we affect the chances of successfully receiving a packet
from either responder due to the variation in power loss and
propagation delay. For each initiator position, we perform 3,000
ranging exchanges with \crng, measuring the packet reception ratio
(\prr) of \response packets along with the resulting ranging
estimates. As a baseline, we also performed 1,000 ranging exchanges
with each responder in isolation, yielding $\prr=100\%$ for all
initiator positions.

\begin{figure}[!tb]
\centering
\includegraphics{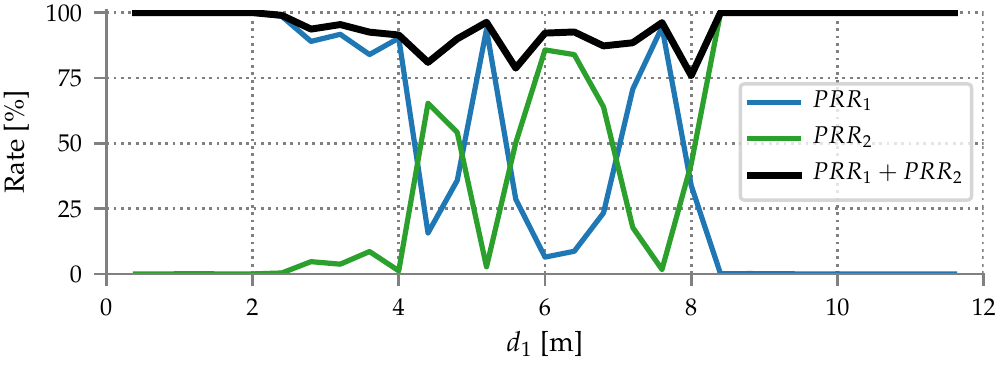}
\caption{Packet reception rate (\prr) \vs initiator position $d_1$, 
  with two concurrent transmissions.}
\label{fig:capture-test-prr}
\end{figure} 

Figure~\ref{fig:capture-test-prr} shows the $\prr_i$ of each responder
and the overall $\aggprr = \prr_{1} + \prr_{2}$ denoting the case in
which a packet from either responder is received correctly. Among all
initiator positions, the worst overall \aggprr~$=$~75.93\% is achieved
for $d_1 = 8$~m. On the other hand, placing the initiator close to one of the
responders (\ie $d_1 \leq 2$~m or $d_1 \geq 10$~m) yields
$\aggprr \geq 99.9\%$. 
We also observe strong fluctuations in the center area. For instance,
placing the initiator at $d_1 = 5.2$~m yields $\prr_{1}= 93.6$\% and
$\prr_{2}=2.7$\%, while nudging it at $d_1 = 6$~m yields 
$\prr_{1}= 6.43$\% and $\prr_{2}=85.73$\%. 

\fakeparagraph{Summary} 
Overall, this experiment confirms the ability of the
DW1000 to successfully decode, with high probability, 
one of the packets from concurrent transmissions.

\subsection{How Concurrent Transmissions Affect Ranging Accuracy?}
\label{sec:obs:accuracy}

We also implicitly assumed that concurrent transmissions do not affect
the ranging accuracy.  In practice, however, the UWB wireless channel
is far from being as ``clean'' as in the idealized view of
Figure~\ref{fig:concurrent-uwb}. The first path is typically followed
by several multipath reflections, which effectively create a ``tail''
after the leading signal. Depending on its temporal and spatial
displacement, this tail may interfere with the first path of other
responders by
\begin{inparaenum}[\itshape i)]
\item reducing its amplitude, or
\item generating MPC that can be mistaken for the first path, 
	inducing estimation errors. 
\end{inparaenum}
Therefore, we now ascertain whether concurrent transmissions degrade
ranging accuracy.

\begin{figure}[!tb]
\centering
\subfloat[Isolated responders.\label{fig:rng-hist}]{
  \includegraphics{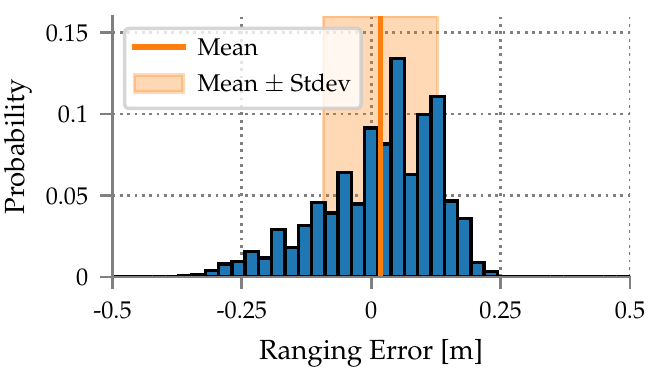}}
\hfill
\subfloat[Concurrent responders.\label{fig:rng-stx-hist}]{
  \includegraphics{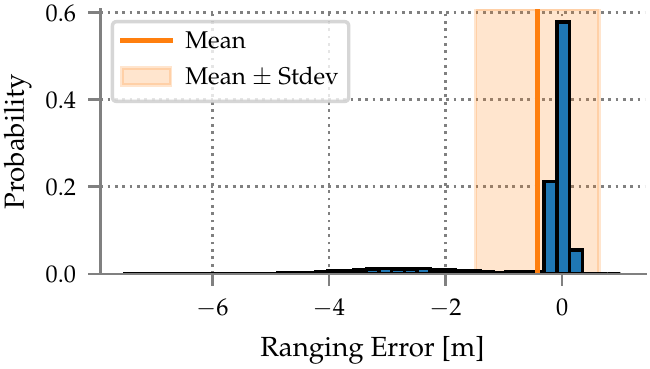}}
\caption{Normalized histogram of the ranging error with responders in 
  isolation (Figure~\ref{fig:rng-hist}) 
  \vs two concurrent responders (Figure~\ref{fig:rng-stx-hist}). 
  In the latter, the initiator sometimes receives 
  the \response from the farthest responder while estimating the 
  first path from the closest one, therefore increasing the absolute error.}
\label{fig:rng-error-hist}
\end{figure}

\fakeparagraph{Baseline: Isolated Responders} We first look at the
ranging accuracy for all initiator positions with each responder
\emph{in isolation}, using the same setup of
Figure~\ref{fig:capture-exp-deployment}. Figure~\ref{fig:rng-hist}
shows the normalized histogram of the resulting ranging error from
58,000 ranging measurements. The average error is $\mu = 1.7$~cm, with
a standard deviation $\sigma=10.9$~cm. The maximum absolute error is
37~cm. The median of the absolute error is 8~cm, while the \nth{99}
percentile is 28~cm. These results are in accordance with previously
reported studies~\cite{surepoint,polypoint} employing the DW1000
transceiver.

\begin{figure}[!tb]
\centering
\subfloat[Ranging Error $\in {[-7.5, -0.5]}$.\label{fig:rng-error-hist-zoom-1}]{
  \includegraphics{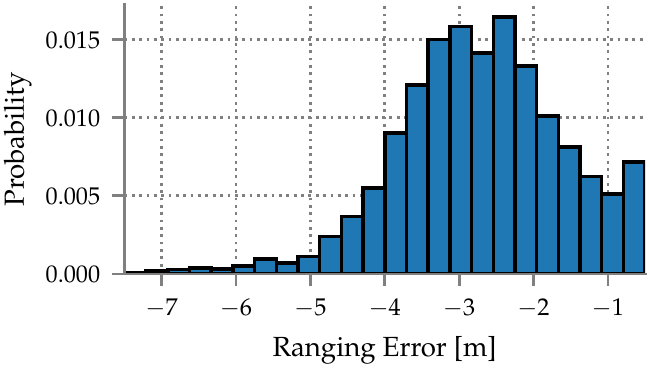}}
\hfill
\subfloat[Ranging Error $\in {[-0.5, 0.5]}$.\label{fig:rng-error-hist-zoom-2}]{
  \includegraphics{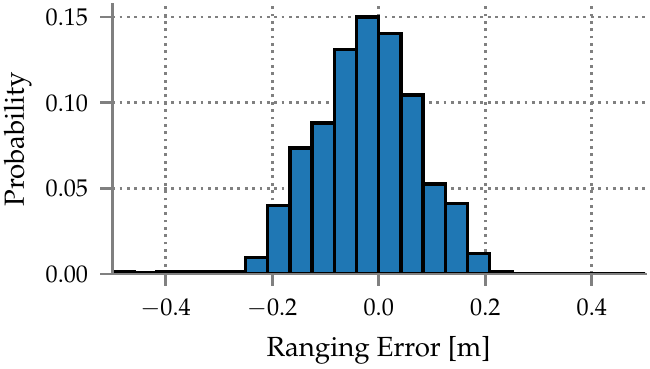}}
\caption{Zoomed-in views of Figure~\ref{fig:rng-stx-hist}.}
\label{fig:rng-error-hist-zoom}
\end{figure}

\fakeparagraph{Concurrent Responders: Impact on Ranging Accuracy}
Figure~\ref{fig:rng-stx-hist} shows the normalized histogram of the
ranging error of 82,519 measurements using instead two concurrent
responders\footnote{Note we do not obtain valid ranging measurements
  in case of RX errors due to collisions.}. The median of
the absolute error is 8~cm, as in the isolated case, while the
\nth{25} and \nth{75} percentiles are 4~cm and 15~cm, respectively.
However, while the average error $\mu = -0.42$~cm is comparable, the
standard deviation $\sigma = 1.05$~m is significantly higher.
Further, the error distribution is clearly different \wrt the case of
isolated responders (Figure~\ref{fig:rng-hist}); to better appreciate
the trends, Figure~\ref{fig:rng-error-hist-zoom} offers a zoomed-in
view of two key areas of the histogram in
Figure~\ref{fig:rng-stx-hist}. Indeed, the latter has a long tail of
measurements with significant errors; for 14.87\% of the measured
samples the ranging error is $<-0.5$~m, while in the isolated case the
maximum absolute error only reaches 37~cm.

\setlength{\columnsep}{8pt}
\setlength{\intextsep}{4pt}
\begin{wrapfigure}{R}{5.6cm}
\centering
\includegraphics{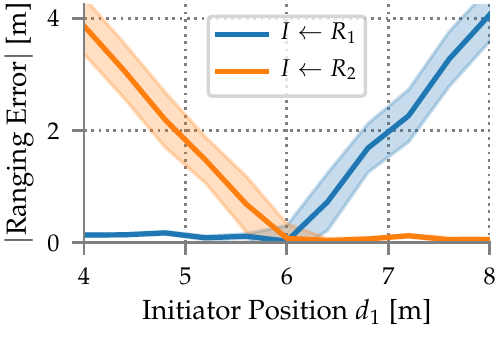}
\caption{Ranging error \vs initiator position.}
\label{fig:stx-rng-error-vs-pos}
\end{wrapfigure}

\fakeparagraph{The Culprit: 
  Mismatch between Received \response and Nearest Responder} 
To understand why, we study the ranging error when the initiator
is located in the center area ($4 \leq d_1 \leq 8$), the one with major \prr
fluctuations (Figure~\ref{fig:capture-test-prr}). 
Figure~\ref{fig:stx-rng-error-vs-pos} shows the average
absolute ranging error of the packets received from each responder as a
function of the initiator position. Colored areas represent the standard
deviation.

The ranging error of $R_1$ and $R_2$ increases dramatically for $d_1 \geq 6$~m
and $d_2 \geq 6$~m, respectively. Moreover, the magnitude of the error exhibits
an interesting phenomenon. For instance, when the initiator is at
$d_1 = 6.8$~m, the average error for \response packets 
received from $R_1$ is 1.68~m, very close to the displacement between responders,
${\deltad = \mid d_1 - d_2 \mid = \mid 6.8 - 5.2\mid = 1.6}$~m. Similarly, for
$d_1 = 5.2$~m and $\deltad = 1.6$~m, the average error for the packets received
from $R_2$ is 1.47~m. 

The observation that the ranging error approximates the displacement
\deltad between responders points to the fact that these high errors
appear when the initiator receives the \response from the farthest
responder but estimates the first path of the signal with the CIR
peak corresponding instead to the nearest responder. This phenomenon
explains the high errors shown in Figure~\ref{fig:rng-stx-hist} 
and~\ref{fig:rng-error-hist-zoom-1},
which are the result of this mismatch between the successful responder
and the origin of the obtained first path. In fact, the higher
probabilities in Figure~\ref{fig:rng-error-hist-zoom-1} 
correspond to positions where the responder farther from the
initiator achieves the highest $\prr_i$ in
Figure~\ref{fig:capture-test-prr}. For example, for $d_1 = 7.6$~m, the
far responder $R_1$ achieves $\prr_1 = 94.46$\% and an average ranging
error of $-3.27$~m, which again corresponds to $\deltad = 3.2$~m and
also to the highest probability in Figure~\ref{fig:rng-error-hist-zoom-1}.

\fakeparagraph{The Role of TX Scheduling Uncertainty} When this
mismatch occurs, we also observe a relatively large standard deviation
in the ranging error. This is generated by the 8~ns TX
scheduling granularity of the DW1000 transceiver (\ref{sec:crng}). In
\sstwr (Figure~\ref{fig:two-way-ranging}), responders insert in the
\response the elapsed time \mbox{$\dtx = \trtx - \tprx$} 
between receiving the \poll and sending the \response. 
The initiator uses \dtx to precisely
estimate the time of flight of the signal. However, the 8~ns
uncertainty produces a discrepancy on \trtx, 
and therefore between the \dtx used by the
initiator and obtained from the successful \response
and the \dtx actually applied by the closest responder, 
resulting in significant error variations.

\fakeparagraph{Summary} 
Concurrent transmissions can negatively affect ranging
by producing a mismatch between the successful responder and the 
detected CIR path used to compute the time of flight. 
However, we also note that 84.59\% of the concurrent ranging samples 
are quite accurate, achieving an absolute error $< 30$~cm. 

\subsection{Does the CIR Contain Enough Information for Ranging?}
\label{sec:cir-enough}

In~\ref{sec:crng} we have mentioned that the limitation on the
granularity of TX scheduling in the DW1000 introduces an 8~ns
uncertainty. Given that an error of 1~ns in estimating the time of
flight results in a corresponding error of $\approx$30~cm, this
raises questions to whether the information in the CIR is sufficient
to recover the timing information necessary for distance estimation.

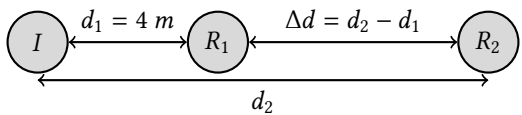
\begin{figure}[!b]
\centering
\begin{tikzpicture}[xscale=0.6]%
\tikzstyle{device} = [circle, thick, draw=black, fill=gray!30, minimum size=8mm]
\node[device] (a) at (0, 0) {$I$};
\node[device] (b) at (4, 0) {$R_1$};
\node[device] (c) at (10, 0) {$R_2$};
\draw[<->, thick] (a) -- (b) node[pos=.5,above]{$d_1 = 4~m$};
\draw[<->, thick] (b) -- (c) node[pos=.5,above]{$\Delta d = d_2 - d_1$};
\draw[<->, thick] (0, -.5) -- (10, -.5) node[pos=.5,below]{$d_2$};
\end{tikzpicture}
\caption{Experimental setup to analyze the CIR resulting from concurrent
  ranging (\ref{sec:cir-enough}).}
\label{fig:chorus-exp-deployment}
\end{figure}

We run another series of experiments using again three nodes but
arranged slightly differently (Figure~\ref{fig:chorus-exp-deployment}). 
We set $I$ and $R_1$ at a fixed distance $d_1 = 4$~m, 
and place $R_2$ at a distance $d_2 > d_1$ from $I$; 
the two responders are therefore separated by a distance
$\deltad = d_2 - d_1$.  Unlike previous experiments, we increase $d_2$
in steps of 0.8~m; we explore $4.8 \leq d_2 \leq 12$~m, and therefore
$0.8 \leq \deltad \leq 8$~m. For each position of $R_2$, we run the
experiment until we successfully receive 500~\response packets,
i.e., valid ranging estimates; we measure the CIR on the
initiator after each received \response.

\begin{figure}[!t]
\centering
\includegraphics{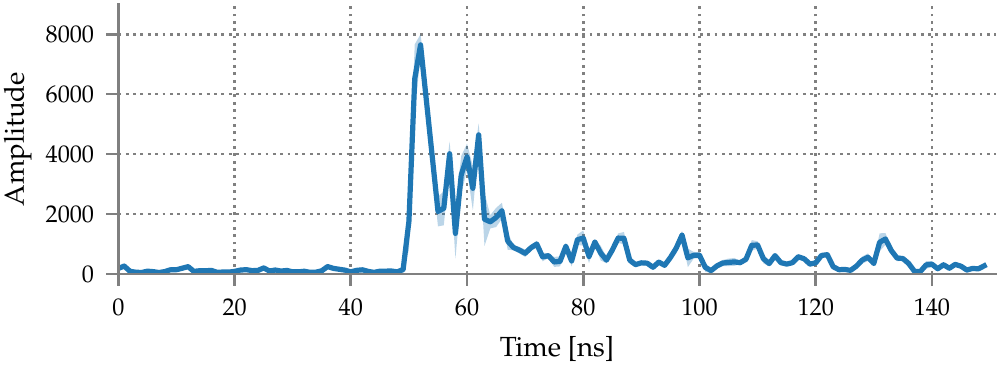}
\caption{Average amplitude and standard deviation of 500~CIR signals for an
  isolated responder at $d_1=4$~m.}
\label{fig:single-tx-cir-variations}

\includegraphics{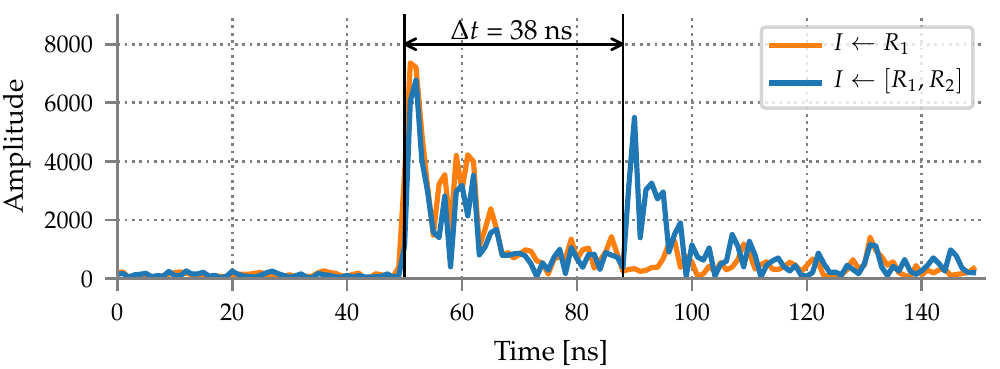}
\caption{Impact of concurrent transmissions on the CIR. The \response TX from
  \resp{2} introduces a second peak at a time shift $\Delta t = 38$~ns after
  the direct path from \resp{1}.}
\label{fig:chorus-cir}

\includegraphics{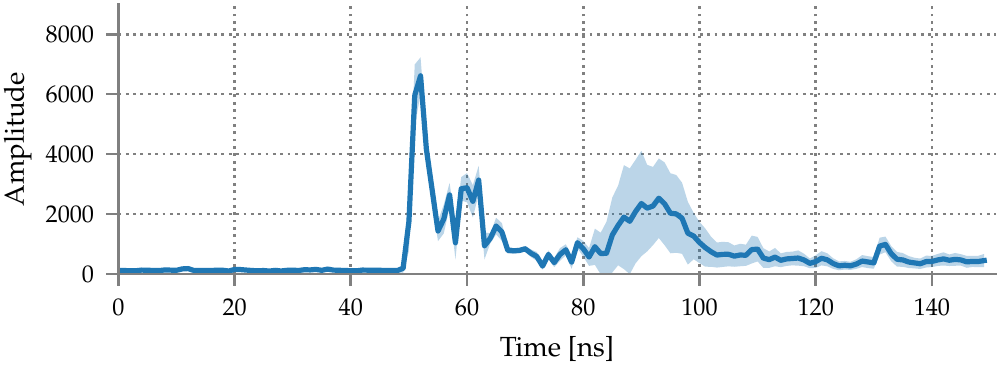}
\caption{Average amplitude and standard deviation of 500 CIR signals, 
  aligned based on the \fpindex, for two concurrent responders 
  at distance ${d_1 = 4}$~m and ${d_2 = 9.6}$~m from the initiator. 
}
\label{fig:chorus-cir-variations}
\end{figure}

\fakeparagraph{Baseline: Isolated Responders} 
Before using concurrent responders, we first measured the CIR of 
$R_1$ ($d_1=4$~m) in isolation. Figure~\ref{fig:single-tx-cir-variations} 
shows the average amplitude and standard deviation across 500~CIR signals, 
averaged by aligning them to the first path index (\fpindex) 
reported by the DW1000~(\ref{sec:dw1000}). 
The measured CIR presents an evident direct path at 50~ns, 
followed by strong multipath. 
We observe that the CIR barely changes across the 500~signals, 
exhibiting only minor variations in these MPCs (around 55--65~ns).

\fakeparagraph{Concurrent Responders: Distance Estimation} We now
analyze the effect of $R_2$ transmitting \textit{concurrently} with
$R_1$, and show how the distance of $R_2$ can be estimated. We focus
on a single distance $d_2 = 9.6$~m and on a single CIR
(Figure~\ref{fig:chorus-cir}), to analyze in depth the phenomena at
stake; we later discuss results acquired from 500~CIR signals
(Figure~\ref{fig:chorus-cir-variations}) and for other $d_2$ values
(Table~\ref{table:concurrent-ranging}).

Figure~\ref{fig:chorus-cir} shows that the \response of $R_2$ introduces a
second peak in the CIR, centered around 90~ns. This is compatible with our
a-priori knowledge of $d_2 = 9.6$~m; the question is whether this distance can
be estimated from the CIR.

Placing the direct path from $R_2$ in time constitutes a
problem per se. In the case of $R_1$, this estimation is performed
accurately and automatically by the DW1000, enabling an accurate
estimate of $d_1$. The same could be performed for $R_2$ if it were in
isolation, but not concurrently with $R_1$. 
Therefore, here we estimate the direct path from $R_2$
as the CIR index whose signal amplitude is closest to $20\%$ of
the maximum amplitude of the peak---a simple technique used, \eg
in~\cite{harmonium}. The offset between the CIR index and the one
returned by the DW1000 for $R_1$, for which a precise estimate is
available, returns the delay \deltat between the responses of $R_1$
and $R_2$. We investigate more sophisticated and accurate
techniques in~\ref{sec:reloaded}.

The value of \deltat is induced by the propagation delay
caused by the difference $\deltad = d_2 - d_1$ in the distance of the
responders from the initiator.  Recall the basics of \sstwr
(\ref{sec:toa}, Figure~\ref{fig:two-way-ranging}) and of concurrent ranging
(\ref{sec:crng}, Figure~\ref{fig:crng}). $R_2$ receives the
\poll from $I$ slightly after $R_1$; the propagation of the \response
back to $I$ incurs the same delay; therefore, the response from $R_2$
arrives at $I$ with a delay $\deltat = 2 \times \frac{\deltad}{c}$
\wrt $R_1$.

In our case, the estimate above from the CIR signal yields
$\deltat=38$~ns, corresponding to 
\mbox{$\deltad \approx 5.6$~m}---indeed the displacement
of the two responders. Therefore, by knowing the distance $d_1$ between
$I$ and $R_1$, estimated precisely by the DW1000, we can easily estimate
the distance between $I$ and $R_2$ as $d_2 = d_1 + \deltad$. 
This confirms that a single concurrent ranging exchange 
contains enough information to reconstruct both distance estimates.

\fakeparagraph{Concurrent Transmissions: Sources of Ranging Error}
Another way to look at Figure~\ref{fig:chorus-cir} is to compare it
against Figure~\ref{fig:concurrent-uwb}; while the latter provides an
\emph{idealized} view of what happens in the UWB channel,
Figure~\ref{fig:chorus-cir} provides a \emph{real} view. 
Multipath propagation and interference among the different paths 
of each signal affects the measured CIR; 
it is therefore interesting to see whether this holds in general 
and what is the impact on the (weaker) signal from $R_2$.

To this end, Figure~\ref{fig:chorus-cir-variations} shows the average
amplitude and standard deviation of 500~CIR signals aligned based on
the \fpindex with $d_1 = 4$~m, and $d_2 = 9.6$~m.  We observe that the
first pulse, the one from the closer $R_1$, presents only
minor variations in the amplitude of the direct path and of MPC,
coherently with Figure~\ref{fig:single-tx-cir-variations}.  In
contrast, the pulse from $R_2$ exhibits stronger variations, as shown
by the colored area between 80 and 110~ns representing the standard
deviation. However, these variations can  be ascribed only marginally
to interference with the pulse from $R_1$; we argue, and provide
evidence next, that these variations are caused by the result of small
time shifts of the observed CIR pulse, in turn caused by the
$\epsilon \in [-8,0)$~ns TX scheduling uncertainty.

\begin{figure}[!t]
\centering
\includegraphics{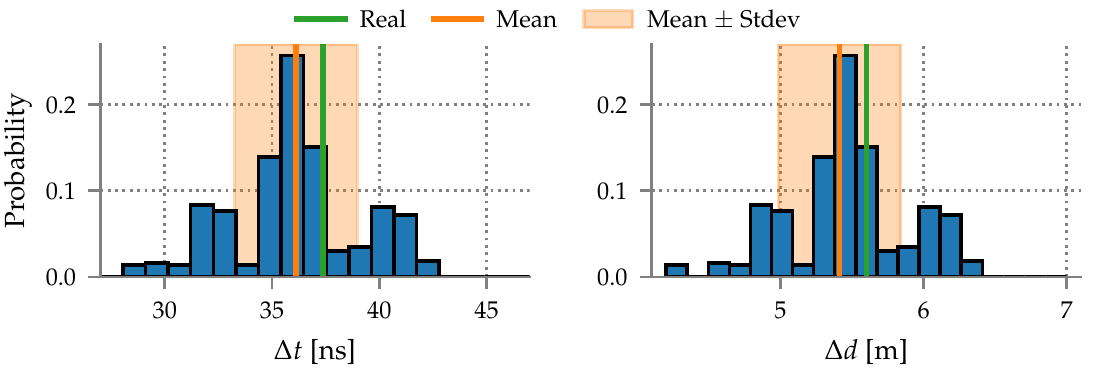}
\caption{Normalized histograms of the time offset \deltat and corresponding
  distance offset \deltad between the leading CIR pulses from $R_1$ and $R_2$.}
\label{fig:chorus-at-hist}
\end{figure}

\fakeparagraph{TX Uncertainty Affects Time Offsets}
Figure~\ref{fig:chorus-at-hist} shows the normalized histogram, for the same
500~CIR signals, of the time offset \deltat between the times at which the
responses from $R_1$ and $R_2$ are received at $I$. The real value, computed
with exact knowledge of distances, is \deltat=~37.37~ns; the average from the
CIR samples is instead \deltat$= 36.11$~ns, with $\sigma = 2.85$~ns. 
These values, and the trends in Figure~\ref{fig:chorus-at-hist}, are
compatible with the 8~ns uncertainty deriving from TX scheduling.

\fakeparagraph{Time Offsets Affect Distance Offsets} As shown in
Figure~\ref{fig:chorus-at-hist}, the uncertainty in time offset
directly translates into uncertainty in the distance offset, whose
real value is $\deltad=~5.6$~m. In contrast, the average estimate is
$\deltad = 5.41$~m, with $\sigma = 0.43$~m. The average error is
therefore $-18$~cm; the \nth{50}, \nth{75}, and \nth{99}
percentiles are 35~cm, 54~cm and 1.25~m, respectively. These results 
still provide sub-meter ranging accuracy as long as the estimated
distance to $R_1$ is accurate enough.

\begin{figure}[!t]
\centering
\includegraphics{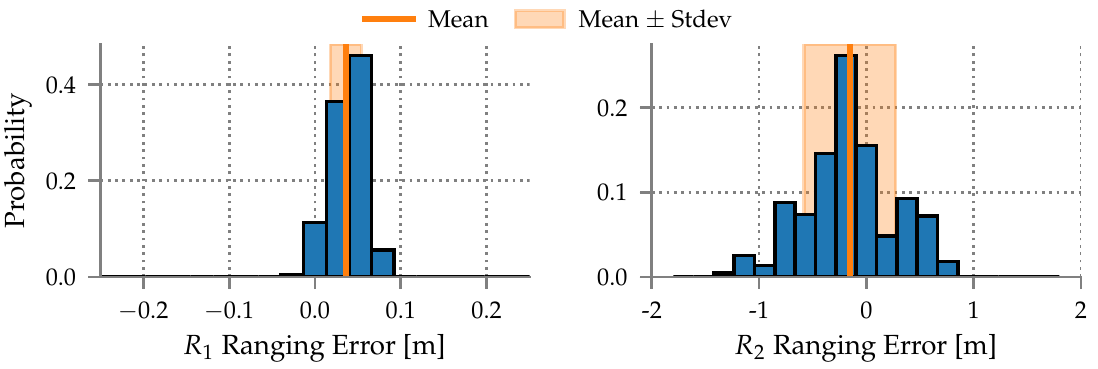}
\caption{Normalized histograms of the concurrent ranging error of each responder.}
\label{fig:concurrent-ranging-error}
\end{figure}

\fakeparagraph{Distance Offsets Affect Ranging Error} Recall that the
distance $d_1$ from $R_1$ to $I$ is obtained directly from the
timestamps provided by the DW1000, while for $R_2$ is estimated as
$d_2 = d_1 + \deltad$.  Therefore, the uncertainty in the distance
offset \deltad directly translates into an additional ranging error,
shown in Figure~\ref{fig:concurrent-ranging-error} for each responder.
$R_1$ exhibits a mean ranging error $\mu = 3.6$~cm with
$\sigma = 1.8$~cm and a \nth{99} percentile over the absolute error of
only 8~cm. Instead, the ranging error for $R_2$, computed indirectly
via \deltad, yields $\mu = -15$~cm with $\sigma = 42.67$~cm. 
The median of the absolute error of $R_2$ is 31~cm, 
while the \nth{25}, \nth{75}, and \nth{99} percentiles 
are 16~cm, 58~cm, and 1.18~m, respectively.

\fakeparagraph{Impact of Distance between Responders} 
In principle, the results above demonstrate the feasibility of 
concurrent ranging and its ability to achieve sub-meter accuracy. 
Nevertheless, these results were obtained for a single value of $d_2$.
Table~\ref{table:concurrent-ranging} summarizes the results 
obtained by varying this distance as described at the beginning of the
section. We only consider the \response packets successfully sent by
$R_1$, since those received from $R_2$ produce the mismatch mentioned
in~\ref{sec:obs:accuracy}, increasing the error by $\approx$~\deltad;
we describe a solution to this latter problem in~\ref{sec:reloaded}.

\begin{table*}[!t]
\centering
\caption{Concurrent ranging performance with two responders \resp{1}
  at a fixed distance $d_1 = 4$~m and \resp{2}
    at different distances $d_2 = d_1 + \Delta d$.}
\label{table:concurrent-ranging}
\resizebox{\textwidth}{!}{
\small
\begin{tabular}{c c | c c c | c c | c c c c c | c c c c c}
    \toprule
    & &
    \multicolumn{3}{c}{\bfseries \prr[\%]} &
    \multicolumn{2}{c}{\bfseries $\mathbf{\Delta d}$ [m]} &
    \multicolumn{5}{c}{\bfseries $\mathbf{R_1}$ Ranging Error [cm]} &
    \multicolumn{5}{c}{\bfseries $\mathbf{R_2}$ Ranging Error [cm]}\\
 	\cmidrule(lr){3-5} \cmidrule(lr){6-7} \cmidrule(lr){8-12} \cmidrule(lr){13-17}
    $\mathbf{d_2}$ & $\mathbf{\Delta d}$ & $PRR_{1}$ & $PRR_{2}$ & $\overline{\prr}$
    & $\mu$ & $\sigma$
    & $\mu$ & $\sigma$ & \nth{50} & \nth{75} & \nth{99}
    & $\mu$ & $\sigma$ & \nth{50} & \nth{75} & \nth{99}\\ 
    \cmidrule(lr){1-17}
	$\mathbf{4.8}$ & $\mathbf{0.8}$ & 2.54 & 95.31 & 97.85 & 0.33 & 0.27 & 3 & 9 
        & 6 & 7 & 26 & -43 & 32 & 30 & 73 & 105\\
	$\mathbf{5.6}$ & $\mathbf{1.6}$ & 36.3 & 36.73 & 73.03 & 1.5 & 0.38 & 6 & 2 
        & 6 & 8 & 12 & -4 & 38 & 31 & 43 & 83\\
	$\mathbf{6.4}$ & $\mathbf{2.4}$ & 65.04 & 22.09 & 87.13 & 2.09 & 0.76 & 6 & 2 
        & 5 & 7 & 10 & -25 & 76 & 51 & 113 & 161\\
	$\mathbf{7.2}$ & $\mathbf{3.2}$ & 0.2 & 99.6 & 99.8 & 3.0 & 0.0 & 8 & 0 & 8 
        & 8 & 8 & -12 & 0 & 12 & 12 & 12\\
	$\mathbf{8.0}$ & $\mathbf{4.0}$ & 38.12 & 44.55 & 82.67 & 4.07 & 0.46 & 8 & 2 
        & 9 & 10 & 13 & 16 & 46 & 41 & 58 & 96\\
	$\mathbf{8.8}$ & $\mathbf{4.8}$ & 69.23 & 20.39 & 89.62 & 4.78 & 0.38 & 5 & 2 
        & 5 & 6 & 9 & 3 & 38 & 25 & 43 & 86\\
	$\mathbf{9.6}$ & $\mathbf{5.6}$ & 100.0 & 0.0 & 100.0 & 5.41 & 0.43 & 4 & 2 & 4 
        & 5 & 8 & -15 & 43 & 31 & 58 & 118\\
	$\mathbf{10.4}$ &$\mathbf{6.4}$ & 94.76 & 2.52 & 97.28 & 6.42 & 0.44 & 5 & 2 
        & 5 & 7 & 9 & 7 & 44 & 36 & 53 & 99\\
	$\mathbf{11.2}$ &$\mathbf{7.2}$ & 85.05 & 5.23 & 90.27 & 7.16 & 0.4 & 6 & 2 
        & 6 & 7 & 10 & 2 & 39 & 34 & 42 & 97\\
	$\mathbf{12.0}$ &$\mathbf{8.0}$ & 100.0 & 0.0 & 100.0 & 8.06 & 0.35 & 4 & 2 
        & 5 & 6 & 9 & 11 & 35 & 29 & 44 & 77\\
	\bottomrule
\end{tabular}
} 
\end{table*}

To automatically detect the direct path of \resp{2}, 
we exploit our a-priori knowledge of where it 
should be located based on \deltad, and therefore \deltat. 
We consider the slice of the CIR defined by $\deltat \pm 8$~ns, 
and detect the first peak in it, 
estimating the direct path as the preceding index with the amplitude 
closest to the $20\%$ of the maximum amplitude, as described earlier. 
To abate false positives, we also enforce the additional constraints 
that a peak has a minimum amplitude of 1,500 and that the minimum distance
between peaks is 8~ns. 

As shown in Table~\ref{table:concurrent-ranging}, the distance to
$R_1$ is estimated with an average error $\mu<9$~cm and
$\sigma < 10$~cm for all tested $d_2$ distances.  The \nth{99}
percentile absolute error is always $< 27$~cm.  These results are in
line with those obtained in~\ref{sec:obs:accuracy}.  As for $R_2$, we
observe that the largest error of the estimated $\Delta d$, and of
$d_2$, is obtained for the shortest distance $d_2 = 4.8$~m.  In this
particular setting, the pulses from both responders are very close and
may even overlap in the CIR, increasing the resulting error,
$\mu=-43$~cm for $d_2$.  The other distances exhibit $\mu\leq 25$~cm.
We observe that the error is significantly lower with
$\Delta d \geq 4$~m, achieving $\nth{75} <60$~cm for $d_2$. Similarly,
for all $\Delta d \geq 4$~m except $\Delta d = 5.6$~m, the \nth{99}
percentile is $< 1$~m.  These results confirm that concurrent ranging
can achieve sub-meter ranging accuracy, as long as the distance
$\Delta d$ between responders is sufficiently large. 

\fakeparagraph{Summary} 
Concurrent ranging can achieve sub-meter accuracy, but requires 
\begin{inparaenum}[\itshape i)]
\item a sufficiently large difference $\Delta d$ in distance (or
  \deltat in time) among concurrent responders, to distinguish the
  responders first paths within the CIR, and
\item a successful receipt of the \response packet from the closest
  responder, otherwise the mismatch of responder identity increases
  the ranging error to $\approx \Delta d$.
\end{inparaenum}

\subsection{What about More Responders?}
\label{sec:cir-multiple}

We conclude the experimental campaign with our strawman implementation
by investigating the impact of more than two concurrent responders,
and their relative distance, on \prr and ranging accuracy.  If
multiple responders are at a similar distance from the initiator,
their pulses are likely to overlap in the CIR, hampering the
discrimination of their direct paths from MPC. Dually, if the distance
between the initiator and the nearest responder is much smaller \wrt
the others, power loss may render the transmissions of farther responders too
faint to be detected at the initiator, due to the interference from
those of the nearest responder.

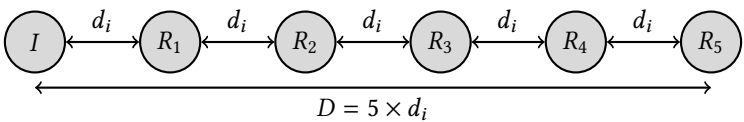
\begin{figure}[!b]
\centering
\begin{tikzpicture}[xscale=0.6]%
\tikzstyle{device} = [circle, thick, draw=black, fill=gray!30, minimum size=8mm]
\node[device] (a) at (0, 0) {$I$};
\node[device] (b) at (3, 0) {$R_1$};
\node[device] (c) at (6, 0) {$R_2$};
\node[device] (d) at (9, 0) {$R_3$};
\node[device] (e) at (12, 0) {$R_4$};
\node[device] (f) at (15, 0) {$R_5$};
\draw[<->, thick] (a) -- (b) node[pos=.5,above]{$d_i$};
\draw[<->, thick] (b) -- (c) node[pos=.5,above]{$d_i$};
\draw[<->, thick] (c) -- (d) node[pos=.5,above]{$d_i$};
\draw[<->, thick] (d) -- (e) node[pos=.5,above]{$d_i$};
\draw[<->, thick] (e) -- (f) node[pos=.5,above]{$d_i$};
\draw[<->, thick] (0, -.6) -- (15, -.6) node[pos=.5,below]{$D = 5 \times d_i$};
\end{tikzpicture}
\caption{Experimental setup to analyze the CIR resulting from five
    concurrent responders (\ref{sec:cir-multiple}).}
\label{fig:five-responders-setup}
\end{figure}

To investigate these aspects, we run experiments with five concurrent
responders arranged in a line (Figure~\ref{fig:five-responders-setup}), 
for which we change the inter-node distance $d_i$. 
For every tested $d_i$, we repeat the experiment until we obtain 
500~successfully received \response packets, as done earlier.

\fakeparagraph{Dense Configuration} We begin by examining a very short
$d_i = 0.4$~m, yielding similar distances between each
responder and the initiator. In this setup, the overall
${\overline{\prr} = 99.36}$\%.

\begin{figure}[!t]
\centering
\subfloat[$d_i = 0.4$~m. The peaks corresponding to each responder are not
  clearly distinguishable; the distance from the initiator cannot be estimated.
  \label{fig:cir-d0-5tx}]{
  \includegraphics{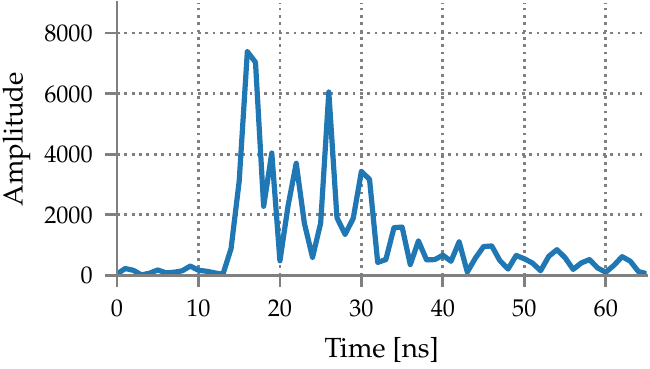}}
\hfill
\subfloat[$d_i = 6$~m. The peaks corresponding to each responder are clearly
  separated; the distance from the initiator can be estimated.\label{fig:cir-d15-5tx}]{
  \includegraphics{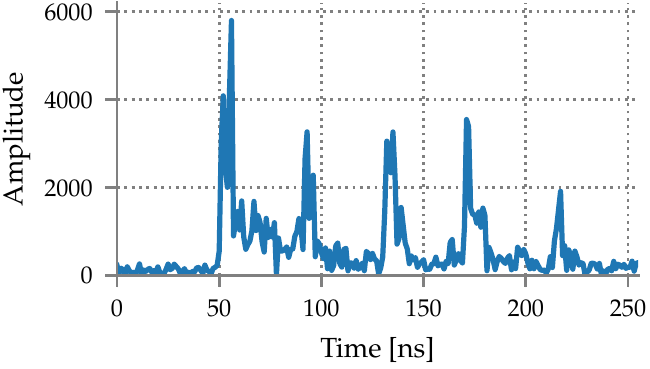}}
\caption{Impact of the relative distance $d_i$ among 5~responders, analyzed
  via the corresponding CIR.}
\end{figure}

Nevertheless, recall that a time-of-flight difference of 1~ns
  translates into a difference of $\approx30$~cm in distance and that
the duration of a UWB pulse is $\leq 2$~ns; pulses from neighboring
responders are therefore likely to overlap, as shown by the CIR in
Figure~\ref{fig:cir-d0-5tx}. Although we can visually observe
different peaks, discriminating the ones associated to responders from
those caused by MPC is very difficult, if not impossible, in absence
of a-priori knowledge about the number of concurrent responders and/or
the environment characteristics. Even when these are present, in some
cases the CIR shows a wider pulse that ``fuses'' the pulses of one or
more responders with MPC. In essence, when the difference in distance
$\deltad = d_i$ among responders is too small, concurrent ranging
cannot be applied with the strawman technique we employed thus far; we
address this problem in~\ref{sec:reloaded}.

\begin{figure}[!t]
\centering
\includegraphics{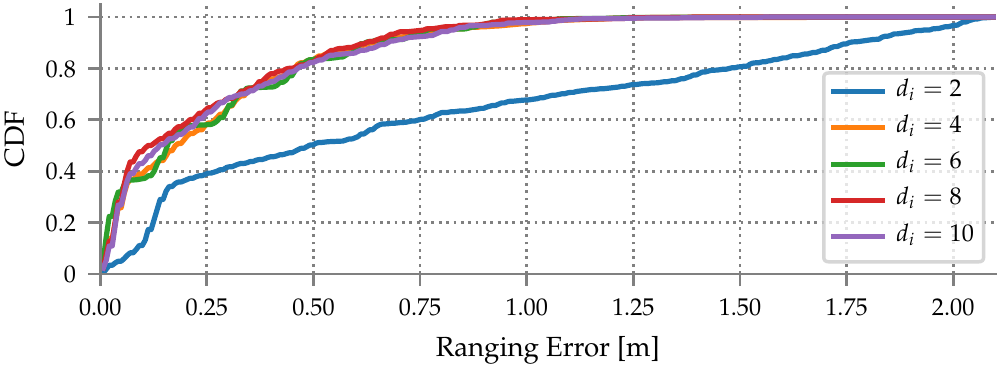}
\caption{Impact of the relative distance $d_i$ among 5 responders: CDF of absolute ranging error.}
\label{fig:crng-ewsn-5tx-cdf}
\end{figure}

\fakeparagraph{Sparser Configurations: \prr} We now explore
${2 \leq d_i \leq 10}$~m, up to a maximum distance $D~=~50$~m 
between the initiator $I$ and the last responder \resp{5}. 
The experiment achieved an overall ${\overline{\prr} = 96.59}$\%, with the minimum $\overline{\prr} = 88.2\%$ for the maximum $d_i = 10$~m, 
and the maximum ${\overline{\prr} = 100}$\%  for $d_i = 8$~m. 
The closest responder $R_1$ achieved \mbox{$\prr_1 =90.56$\%}. 
The \prr of the experiment is interestingly
high, considering that in narrowband technologies
increasing the number of concurrent transmitters sending different packets
typically decreases reliability due to the nature of the
capture effect~\cite{chaos,crystal}. In general, the
  behavior of concurrent transmissions in UWB is slightly
  different---and richer---than in narrowband; the interested reader
  can find a more exhaustive treatment in~\cite{uwb-ctx-fire}. In this
specific case, the reason for the high \prr we observed is the closer
distance to the initiator of $R_1$ \wrt the other responders.

\fakeparagraph{Sparser Configurations: Ranging Error}
Figure~\ref{fig:crng-ewsn-5tx-cdf} shows the CDF of the ranging error
for all distances and responders. We use the same technique
of~\ref{sec:cir-enough} to detect the direct paths and, similarly,
only consider the exchanges (about 90\% in this case) where the
successfully received \response is from the nearest responder
$R_1$, to avoid a mismatch (\ref{sec:obs:accuracy}).

We observe the worst performance for $d_i = 2$~m; peaks from different
responders are still relatively close to each other and affected by
the MPC of previously transmitted pulses. Instead,
Figure~\ref{fig:cir-d15-5tx} shows an example CIR for $d_i = 6$~m, the
intermediate value in the distance range considered. Five distinct
peaks are clearly visible, enabling the initiator to estimate the
distance to each responder. The time offset \deltat between two
consecutive peaks is similar, as expected, given the same distance
offset $\deltad = d_i$ between two neighboring responders. This yields
good sub-meter ranging accuracy for all $d_i\geq 4$, for which 
the average error is $\mu\leq 40$~cm and the absolute error 
  $\nth{75}\leq 60$~cm.

\fakeparagraph{Summary} These results confirm that sub-meter
concurrent ranging is feasible even with multiple responders. However,
ranging accuracy is significantly affected by the relative distance
between responders, which limits practical applicability.

\section{Concurrent Ranging Reloaded}
\label{sec:reloaded}

The experimental campaign in the previous section confirms that
concurrent ranging is feasible, but also highlights several challenges
not tackled by the strawman implementation outlined in~\ref{sec:crng},
limiting the potential and applicability of our technique. In this
section, we overcome these limitations with a novel design
that, informed by the findings in~\ref{sec:questions}, significantly
improves the performance of \crng both in terms of accuracy and
reliability, bringing it in line with conventional methods but at a
fraction of the network and energy overhead.

We begin by removing the main source of inaccuracy, \ie the 8~ns
uncertainty in TX scheduling. The technique we present
(\ref{sec:crng-tosn:txfix}) not only achieves sub-ns precision, as 
shown in our evaluation (\ref{sec:crng-tosn:exp-tx-comp}), but also
doubles as a mechanism to reduce the impact of clock drift, the main
source of error in \sstwr (\ref{sec:soa-sstwr}). We then
present our technique to correctly associate responders with paths
in the CIR (\ref{sec:crng-tosn:resp-id}), followed by two necessary
CIR pre-processing techniques to discriminate the direct paths from MPC 
and noise (\ref{sec:crng-tosn:cir-proc}). Finally, we illustrate two algorithms
for estimating the actual \toa of the direct paths and outline the overall
processing that, by combining all these steps, yields the final
distance estimation (\ref{sec:crng-tosn:time-dist}).
 
\subsection{Locally Compensating for TX Scheduling Uncertainty}
\label{sec:crng-tosn:txfix}

The DW1000 transceiver can schedule a TX in the future with a
precision of $4 / (499.2\times10^6)\approx 8$~ns, much less than the
signal timestamping resolution. \sstwr responders circumvent this lack
of precision by embedding the necessary TX/RX timestamps in their
\response. This is not possible in \crng, and an uncertainty
$\epsilon$ from a uniform distribution $U[-8, 0)$~ns directly affects
concurrent transmissions from responders. The empirical observations
in~\ref{sec:questions} show that mitigating this TX uncertainty is
crucial to enhance accuracy. This section illustrates a
technique, inspired by \decawave engineers during private
communication, that achieves this goal effectively.

A key observation is that both the accurate desired TX timestamp and
the inaccurate one actually used by the radio are \emph{known} at the
responder. Indeed, the DW1000 obtains the latter from the former by
simply discarding its less significant 9~bits.
Therefore, given that the responder knows beforehand the TX timing
error that will occur, it can \emph{compensate} for it while preparing
its \response.

We achieve this by fine-tuning the frequency of the oscillator, an
operation that can be performed entirely in firmware and
\emph{locally} to the responder. In the technique described here, the
compensation relies on the ability of the DW1000 to \emph{trim} its
crystal oscillator frequency~\cite[p.~197]{dw1000-manual-v218} during
operation. The parameter accessible via firmware is the radio
\emph{trim index}, whose value determines the correction currently
applied to the crystal oscillator. By modifying the index by a given
negative or positive amount (\emph{trim step}) we can
increase or decrease the oscillator frequency (\ie clock
speed) and compensate for the aforementioned known TX timing
error. Interestingly, this technique can also be exploited to reduce
the relative \emph{carrier frequency offset} (CFO) between transmitter and
receiver, with the effect of increasing receiver sensitivity,
enhancing CIR estimation, and ultimately improving ranging 
accuracy and precision.

\fakeparagraph{Trim Step Characterization} To design a compensation
strategy, it is necessary to first characterize the impact of a trim
step. To this end, we ran several experiments with a transmitter and a
set of 6~receivers, to assess the impact on the CFO.  The transmitter
is initially configured with a trim index of~0, the minimum allowed,
and sends a packet every 10~ms.  After each TX, a trim step of~+1 is
applied, gradually increasing the index until~31, the maximum allowed,
after which the minimum index of 0 is re-applied; increasing the trim index
reduces the crystal frequency. Receivers do not apply a trim step;
they use a fixed index of~15. For each received packet, we read the
CFO between the transmitter and the corresponding receiver from the
DW1000, which stores in the \texttt{DRX\_CONF} register the receiver
carrier integrator value~\cite[p.~80--81]{dw1000-sw-api} measured
during RX, and convert this value first to Hz and then to
parts-per-million (ppm).

\begin{figure}[!t]
\centering
\includegraphics{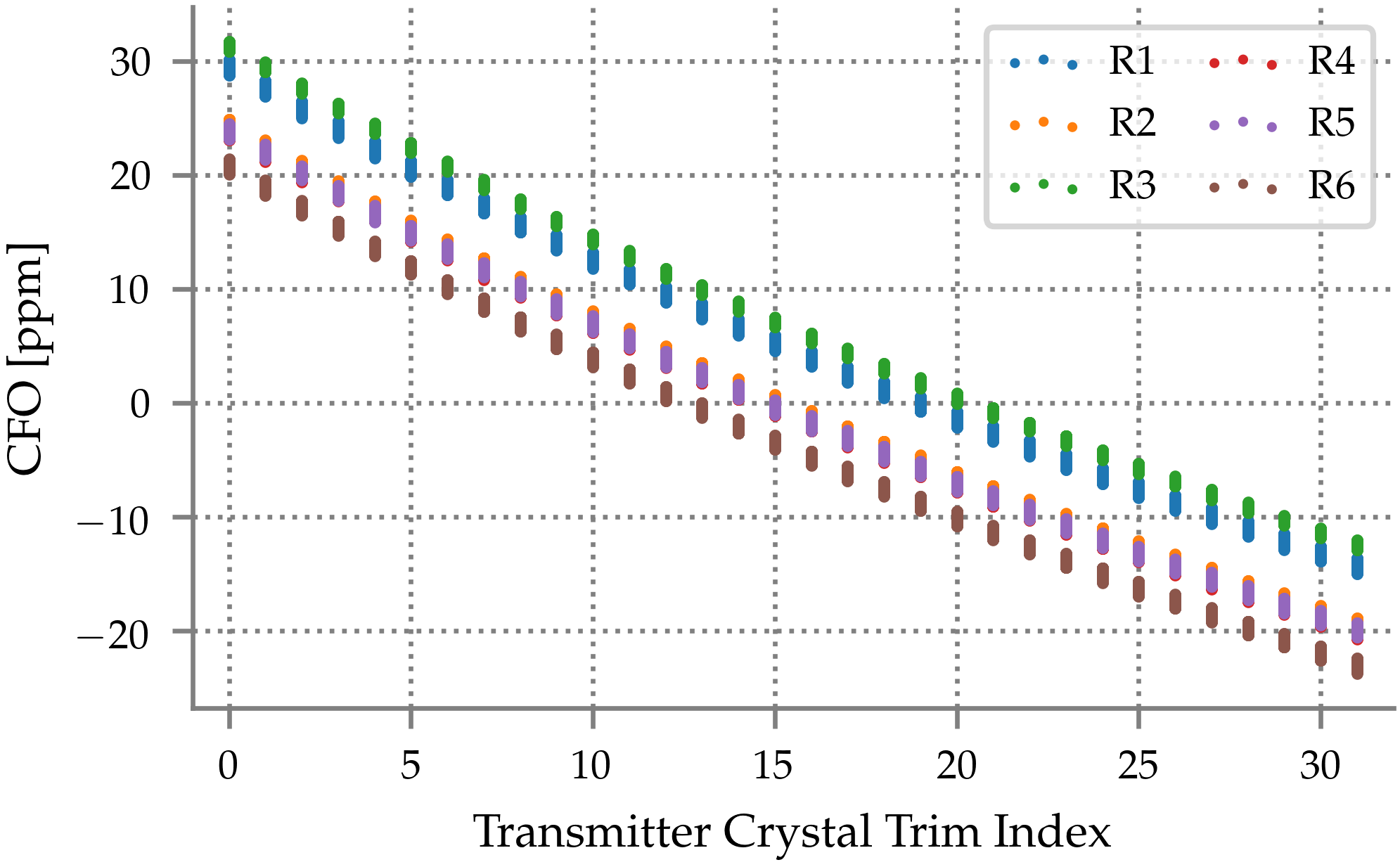}   
\caption{CFO between a transmitter and a set of six receivers, 
  as a function of the transmitter trim index.}
\label{fig:ppm-offset}
\end{figure}

Figure~\ref{fig:ppm-offset} shows the CFO measured for each receiver
as a function of the transmitter trim index, over $\geq$100,000
packets. If the CFO is positive (negative), the receiver local clock
is slower (faster) than the transmitter
clock~\cite[p.~81]{dw1000-sw-api}. All receivers exhibit a
quasi-linear trend, albeit with a different offset. Across many
experiments, we found that the average curve slope is
$\approx -1.48$~ppm per unit trim step. This knowledge is crucial to
properly trim the clock of the responders to match the frequency of
the initiator and compensate for TX uncertainty, as described next.

\fakeparagraph{CFO Adjustment} After receiving the broadcast \poll,
responders obtain the CFO from their carrier integrator and trim their
clock to better match the frequency of the initiator. For instance, if
a given responder measures a CFO of $+3$~ppm, this means that its
clock is slower than the initiator, and its frequency must be
increased by applying a trim step of
$-\frac{\SI{3}{ppm}}{\SI{1.48}{ppm}} \approx -2$.  Repeating this
adjustment limits at $\leq 1$~ppm the absolute value of the CFO
between initiator and responders, reducing the impact of clock drift
and improving RX sensitivity. Moreover, it also improves CIR
estimation, enabling the initiator to better discern the signals from
multiple, concurrent responders and estimate their \toa more
accurately. Finally, this technique can be used to \emph{detune} the
clock (\ie alter its speed), key to compensating for TX uncertainty.

\fakeparagraph{TX Uncertainty Compensation} The DW1000 measures TX and
RX times at the \rmarker (\ref{sec:dw1000}) with \mbox{40-bit}
timestamps in radio time units of $\approx 15.65$~ps.  However, 
when scheduling transmissions, it ignores the lowest 9~bits of the 
desired TX timestamp.
The \emph{known} 9 bits ignored directly inform us of the TX error
$\epsilon \in [-8, 0)$~ns to be compensated for. The compensation
occurs by \emph{temporarily} altering the clock frequency via the trim
index only for a given \emph{detuning interval}, at the end of which
the previous index is restored. Based on the known error $\epsilon$
and the predefined detuning interval $\detuning$, we can easily
compute the trim step
\mbox{$\trimstep = \lfloor \frac{\epsilon}{\SI{1.48}{ppm}\times
    \detuning}\rceil$} to be applied to compensate for the TX
scheduling error.  For instance, assume that a responder knows that
its TX will be anticipated by an error $\epsilon=-5$~ns; its clock
must be slowed down. Assuming a configured detuning interval
$\detuning=\SI{400}{\micro\second}$, a trim step
\mbox{
  $\trimstep = \lfloor \frac{\SI{5}{ns}}{\SI{1.48}{ppm} \times
    \SI{400}{\micro\second}} \rceil = \lfloor 8.45 \rceil = 8$ } must
be applied through the entire interval \detuning.  The rounding,
necessary to map the result on the available integer values of the trim index,
translates into a residual TX scheduling error. This can be easily
removed, after the trim step \trimstep is determined, by recomputing
the detuning interval as
\mbox{$\detuning = \frac{\epsilon}{\SI{1.48}{ppm} \times \trimstep}$},
equal to \SI{422.3}{\micro\second} in our example. Indeed, the
duration of \detuning can be easily controlled in firmware and with a
significantly higher resolution than the trim index, yielding a more
accurate compensation.

\fakeparagraph{Implementation} In our prototype, we determine the trim
step \trimstep, adjust the CFO, and compensate the TX scheduling error
in a single operation. While detuning the clock, we set the data
payload and carry out the other operations necessary before TX,
followed by an idle loop until the detuning interval is over. We then
restore the trim index to the value determined during CFO adjustment
and configure the DW1000 to transmit the \response at the desired
timestamp.  To compensate for an error $\epsilon \in [-8, 0)$~ns
without a very large trim step (\ie abrupt changes of the trim
index) we set a default detuning interval
$\detuning=\SI{560}{\micro\second}$ and increase the ranging response
delay to $\dtx = \SI{800}{\micro\second}$. This value is higher than
the one ($\dtx=\SI{330}{\micro\second}$) used in~\ref{sec:questions}
and, in general, would yield worse \sstwr ranging accuracy due to a
larger clock drift (\ref{sec:toa}). Nevertheless, here we directly
limit the impact of the clock drift with the CFO adjustment, precisely
scheduling transmissions with $<1$~ns errors, as shown
in~\ref{sec:crng-tosn:exp-tx-comp}; therefore, in practice,
  the minor increase in \dtx bears little to no impact.

\subsection{Response Identification}
\label{sec:crng-tosn:resp-id}

As observed in~\ref{sec:questions}, if the distance between the
initiator and the responders is similar, their paths and MPC overlap
in the CIR, hindering responder identification and \toa estimation.
Previous work~\cite{crng-graz} proposed to assign a different pulse
shape to each responder and then use a matched filter to associate
paths with responders. However, this leads to responder
mis-identifications, as we showed in~\cite{chorus}, because 
the channel cannot always be assumed to be separable, \ie the measured peaks in
the CIR can be a combination of multiple paths, and the
received pulse shapes can be deformed, creating ambiguity in the
matched filter output.

To reliably separate and identify responders, we resort to response
position modulation~\cite{crng-graz}, whose effectiveness has instead
been shown by our own work on Chorus~\cite{chorus} and by
SnapLoc~\cite{snaploc}. The technique consists of delaying each
\response by \mbox{$\delta_i = (i - 1)\tdelay$}, where $i \in \{1,N\}$
is the responder identifier. The resulting CIR consists of an ordered
sequence of signals that are time-shifted based
on \begin{inparaenum}[\itshape i)]
\item the assigned delays $\delta_i$, and
\item the propagation delays $\tau_i$,
\end{inparaenum}
as shown in Figure~\ref{fig:crng-tosn:cir-arrangement}.

\begin{figure}[!t]
\centering
\includegraphics{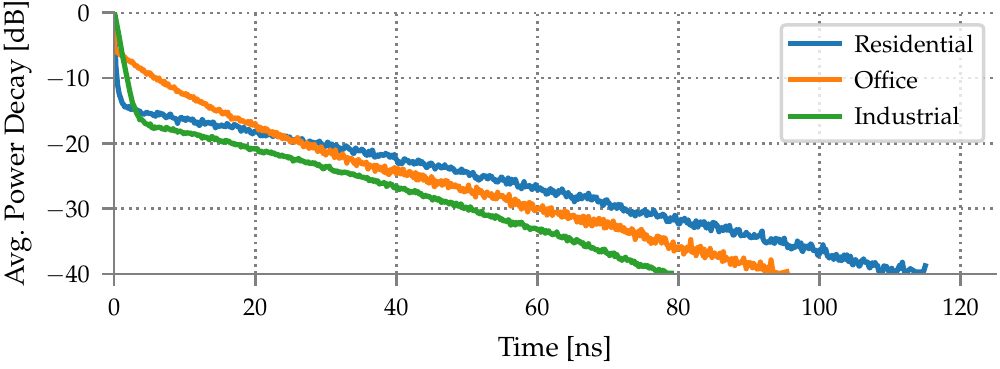}   
\caption{Power decay profile in different environments according to the
{\ieeestd}a radio model~\cite{molisch-ieee-uwb-model}.}
\label{fig:uwb-pdp}
\end{figure}

The constant \tdelay must be set according to
\begin{inparaenum}[\itshape i)]
\item the CIR time span,
\item the maximum propagation time, as determined by the dimensions of
  the target deployment area, and
\item the multipath profile in it.
\end{inparaenum}
Figure~\ref{fig:uwb-pdp} shows the typical power decay profile in
three different environments obtained from the {\ieeestd}a radio
model~\cite{molisch-ieee-uwb-model}.  MPC with a time shift
$\geq 60$~ns suffer from significant power decay \wrt the direct
path. Therefore, by setting $\tdelay = 128$~ns as
in~\cite{chorus,snaploc} we are unlikely to suffer from significant
MPC and able to reliably distinguish the responses. Moreover,
considering that the DW1000 CIR has a maximum time span of
$~\approx 1017$~ns, we can accommodate up to 7~responders, leaving a
small portion of the CIR with only noise.
We observe that this technique relies on the correct
identification of the first and last responder to properly
reconstruct the sequence, and avoid mis-identifications; 
our evaluation (\ref{sec:crng-tosn:eval}) shows that these rarely
occur in practice.

Finally, although the technique is similar to the one
in~\cite{chorus,snaploc}, the different context in which it is applied
yields significant differences. In these systems, the time of flight
$\tau_i$ is \emph{known} and compensated for, based on the fixed and
known position of anchors.  
In \crng, not only $\tau_i$ is not known a
priori, but it also has a twofold impact on the \response RX
timestamps, making the problem more challenging. On the other hand,
\crng is more flexible as it does not rely on the known position of
anchors. Further, as packet exchanges are triggered by the initiator
rather than the anchors as in~\cite{chorus,snaploc}, the former
could determine time shifts on a per-exchange basis, assigning a
different $\delta_i$ to each responder via the broadcast \poll. For
instance, in a case where responders $R_i$ and $R_{i+1}$ have a
distance $d_i \gg d_{i+1}$ from the initiator, a larger time shift
$\delta_{i+1}$ could help separating the pulse of $R_{i+1}$ from the
MPC of $R_i$. Similarly, when more responders are present than what
can be accommodated in the CIR, the initiator could dynamically
determine the responders that should reply and the delays $\delta_i$
they should apply.  This adaptive, initiator-based time shift
assignment opens several intriguing opportunities, especially for
mobile, ranging-only applications; we are currently investigating them
as part of our ongoing work (\ref{sec:discussion}).

\subsection{CIR Pre-processing}
\label{sec:crng-tosn:cir-proc}

We detail two techniques to reorder the CIR array and estimate the
signal noise standard deviation \stdnoise.  These extend and
significantly enhance the techniques we originally proposed 
in~\cite{chorus}, improving the robustness and accuracy 
of the \toa estimation algorithms in~\ref{sec:crng-tosn:toa-est}.

\subsubsection{CIR Array Re-arrangement}
\label{sec:crng-tosn:cir-rearrangement}

\begin{figure}[!t]
\centering
\subfloat[Raw CIR array.\label{fig:crng-tosn:cir-arrangement-raw}]{
  \includegraphics{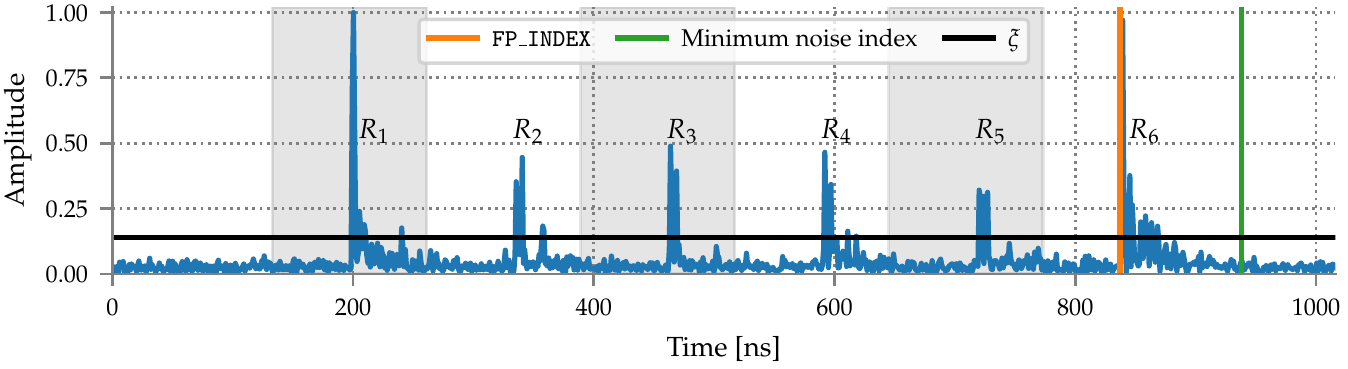}
}\\
\subfloat[Re-arranged CIR array.\label{fig:crng-tosn:cir-arrangement-sorted}]{
 \includegraphics{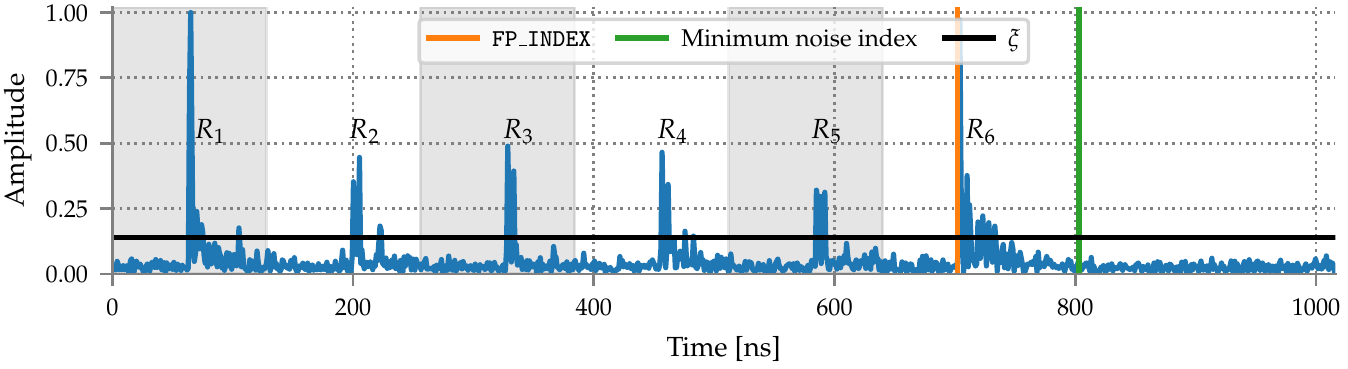}
}
\caption{CIR re-arrangement. The DW1000 measured the \fpindex as the
  direct path of \resp{6} in the raw CIR 
  (Figure~\ref{fig:crng-tosn:cir-arrangement-raw}). 
  After finding the CIR sub-array with the lowest noise, 
  we re-arrange the CIR (Figure~\ref{fig:crng-tosn:cir-arrangement-sorted})
  setting the response of \resp{1} at the beginning and the noise-only
  sub-array at the end.}
\label{fig:crng-tosn:cir-arrangement}
\end{figure}

In the conventional case of an isolated transmitter, the DW1000
arranges the CIR signal by placing the first path at \mbox{\fpindex
  $\approx750$} in the accumulator buffer (\ref{sec:dw1000}).  In
\crng, one would expect the \fpindex to similarly indicate the direct
path of the first responder \resp{1}, \ie the one with the shorter
time shift $\delta_1 = 0$.  Unfortunately, this is not necessarily the
case, as the \fpindex can be associated with the direct path of
\emph{any} of the involved responders
(Figure~\ref{fig:crng-tosn:cir-arrangement}).
Further, and worse, due to the TX time shifts $\delta_i$ we apply in
\crng, the paths associated to the later responders may be circularly
shifted at the beginning of the array, disrupting the implicit
temporal ordering at the core of 
responder identification (\ref{sec:crng-tosn:resp-id}).

Therefore, before estimating the \toa of the concurrent signals, we must
\begin{inparaenum}[\itshape i)]
\item re-arrange the CIR array to match the order expected
  from the assigned time shifts, and
\item correspondingly re-assign the index associated to the \fpindex 
  and whose timestamp is available in radio time units.
\end{inparaenum}
In~\cite{chorus} we addressed a similar problem by 
partially relying on knowledge of the responder ID contained in the 
\response payload (among the several concurrent ones) actually 
decoded by the radio, which then usually places its
corresponding first path at \mbox{\fpindex $\approx750$} in the
CIR. However, this technique relies on successfully decoding a
\response, which is unreliable as we previously observed in~\ref{sec:questions}. 
Here, we remove this dependency and enable a correct CIR re-arrangement 
\emph{even in cases where the initiator is unable to successfully decode
  any \response}, significantly improving reliability.

We achieve this goal by identifying the portion of
the CIR that contains \emph{only} noise, which appears in
between the peaks of the last and first responders. First, 
we normalize the CIR \wrt its maximum amplitude sample and search
for the CIR sub-array of length $W$ with the lowest sum---the
aforementioned noise-only portion. Next, we determine the index at
which this noise sub-array begins (minimum noise index in
Figure~\ref{fig:crng-tosn:cir-arrangement}) and search for the next
sample index whose amplitude is above a threshold $\xi$. This latter
index is a rough estimate of the direct path of \resp{1}, 
the first expected responder. We then re-order the CIR array 
by applying a circular shift, setting the $N$ responses 
at the beginning of the array, followed by the noise-only portion at the end. 
Finally, we re-assign the index corresponding to the original 
\fpindex measured by the DW1000 and whose radio timestamp is available.

We empirically found, by analyzing 10,000s of CIRs signals, that a
threshold $\xi \in [0.12, 0.2]$ yields an accurate CIR reordering. 
Lower values may cause errors due to noise or MPC, while higher
values may disregard a weak first path from \resp{1}. The noise window
$W$ must be set based on the CIR time span, the time shifts $\delta_i$
applied, and the number $N$ of concurrent responders. Hereafter, we
set $\xi = 0.14$ and $W = 228$ samples with $N = 6$ responders and
$\tdelay = 128$~ns.

\subsubsection{Estimating the Noise Standard Deviation}
\label{sec:crng-tosn:noise-std}

\toa estimation algorithms frequently rely on a threshold 
derived from the noise standard deviation \stdnoise, 
to detect the first path from noise and MPC. 
The DW1000 estimates $\stdnoise^{DW}$ based on the measured 
CIR~\cite{dw1000-manual-v218}. However, 
in the presence of concurrent transmissions, the DW1000 sometimes
yields a significantly overestimated $\stdnoise^{DW}$,
likely because it considers the additional \response signals as noise.
Therefore, we recompute our own estimate of \stdnoise 
as the standard deviation of the last 128~samples of the re-arranged CIR
(Figure~\ref{fig:crng-tosn:cir-arrangement-sorted}). By design
(\ref{sec:crng-tosn:cir-rearrangement}) these samples belong to the
noise-only portion at the end of the re-arranged CIR, free from MPC from
responses; the noise estimate is therefore significantly more reliable
than the one computed by the DW1000, meant for non-concurrent
ranging.

\begingroup
\setlength{\columnsep}{8pt}
\setlength{\intextsep}{4pt}
\begin{wrapfigure}{R}{5.6cm}
\centering
\includegraphics{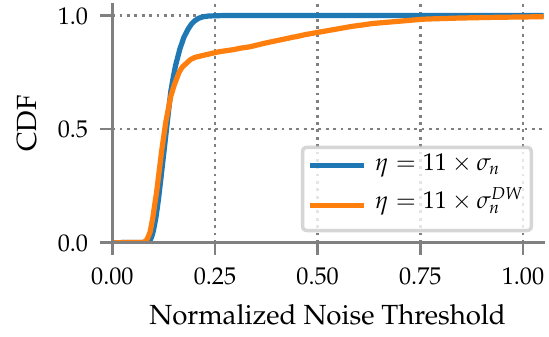}
\caption{Threshold comparison.}
\label{fig:std-noise-cdf}
\end{wrapfigure}

Figure~\ref{fig:std-noise-cdf} offers evidence of this last statement
by comparing the two techniques across the 9,000 signals 
with $N = 6$ concurrent responders we use in~\ref{sec:crng-exp-static} 
to evaluate the performance of \crng with the initiator placed
in 18 different static positions. 
The chart shows the actual noise threshold
computed as $\threshold = 11 \times \stdnoise$, which we empirically
found to be a good compromise for \toa estimation
(\ref{sec:crng-tosn:toa-est}). Using our technique, \threshold
converges to a $\nth{99}$ percentile of $0.213$ over the normalized
CIR amplitude, while the default $\stdnoise^{DW}$ yields
$\nth{99} = 0.921$; this value would lead to discard most of
the peaks from concurrent responders.
For instance, in Figure~\ref{fig:crng-tosn:cir-arrangement} 
only 2 out of 6 direct paths would be detected with such a high threshold.
Across these 9,000 signals, using our estimated $\stdnoise$ 
instead of $\stdnoise^{DW}$ increases the ranging and localization 
reliability of concurrent ranging by up to $16\%$ depending on the
\toa algorithm used, as we explain next.

\subsection{From Time to Distance}
\label{sec:crng-tosn:time-dist}

Enabling \crng on the DW1000 requires a dedicated algorithm
(\ref{sec:crng-tosn:toa-est}) to estimate the \toa of each \response
in the CIR. This timing information must then be translated into the
corresponding distances (\ref{sec:crng-tosn:dist-est}), used directly
or in the computation of the initiator position
(\ref{sec:soa:toa-loc}).

\subsubsection{Time of Arrival Estimation}
\label{sec:crng-tosn:toa-est}

To determine the first path of each responder, we use FFT to upsample
the re-arranged CIR signals by a factor $L = 30$, yielding a
resolution $T_s \approx 33.38$~ps. We then split the CIR into chunks
of length equal to the time shift \tdelay used for responder
identification~(\ref{sec:crng-tosn:resp-id}), therefore effectively
separating the signals of each \response. Finally, the actual \toa
estimation algorithm is applied to each chunk, yielding the CIR index
\indexi marking the \toa of each responder \resp{i}. We consider two
\toa estimation algorithms:

\begin{itemize}
\item \emph{Threshold-based.}  This commonly-used algorithm simply
  places the first path at the first $i^\mathit{th}$ index whose sampled
  amplitude $A_i > \threshold$, where \threshold is the noise
  threshold (\ref{sec:crng-tosn:cir-proc}).

\item \emph{Search and Subtract (\ssub).} This well-known algorithm
  has been proposed in~\cite{dardari-toa-estimation}; here, we use our
  adaptation~\cite{chorus} to the case of concurrent
  transmissions\footnote{Hereafter, we refer to this adaptation simply as
    \ssub, for brevity.}. \ssub determines the $K$ strongest
  paths, considering \emph{all} signal paths whose peak
  amplitude $A_i > \threshold$. The first path is then 
  estimated as the one with the minimum
  time delay, \ie minimum index in the CIR chunk.

\end{itemize}

These two algorithms strike different trade-offs \wrt complexity,
accuracy, and resilience to multipath. The threshold-based algorithm is very
simple and efficient but also sensitive to high noise. For instance,
if a late MPC from a previous chunk appears at the beginning of the next
one with above-threshold amplitude, it is selected as the first
path, yielding an incorrect \toa estimate.  \ssub is more resilient,
as these late MPC from previous responses would need to be stronger
than the $K$ strongest paths from the current chunk to cause a mismatch. 
Still, when several strong MPC are in the same chunk, \ssub may incorrectly select
one of them as the first path, especially if the latter is weaker than
MPC.  Moreover, \ssub relies on a matched filter, which
\begin{inparaenum}[\itshape i)]
\item requires to determine the filter template by measuring the shape
  of the transmitted UWB pulses, and 
\item increases computational complexity, as $K$ discrete convolutions
  must be performed to find the $K$ strongest paths.
\end{inparaenum}

We compare these \toa estimation algorithms in our evaluation
(\ref{sec:crng-tosn:eval}).

\subsubsection{Distance Estimation}
\label{sec:crng-tosn:dist-est}

These \toa estimation algorithms determine the CIR indexes \indexi
marking the direct path of each \response. These, however, are only
\emph{array indexes}; each must be translated into a radio timestamp
marking the \emph{time of arrival} of the corresponding \response, and
combined with other timing information to reconstruct the distance
$d_i$ between initiator and responder. 

In~\ref{sec:crng}--\ref{sec:questions} we relied on the fact that the
radio \emph{directly} estimates the \toa of the first responder
\resp{1} with high accuracy, enabling accurate distance estimation by
using the timestamps embedded in the payload. Then, by looking at the
time difference $\Delta t_{i, 1}$ between the first path
of \resp{1} and another responder $R_i$ we can determine its distance
from the initiator as $d_i = d_1 + c\frac{\Delta t_{i, 1}}{2}$.  This
approach assumes that the radio
\begin{inparaenum}[\itshape i)]
\item places the direct path of \resp{1} at the \fpindex and
\item successfully decodes the \response from \resp{1},
  containing the necessary timestamps to accurately determine $d_1$.
\end{inparaenum}
However, the former is not necessarily true
(Figure~\ref{fig:crng-tosn:cir-arrangement}); as for the latter, the
radio may receive the \response packet from any responder or
none. Therefore, we cannot rely on the distance estimation of
\resp{1}.

Interestingly, the compensation technique to eliminate the TX
scheduling uncertainty (\ref{sec:crng-tosn:txfix}) is also key to
enable an alternate approach avoiding these issues and yielding
additional benefits. Indeed, this technique enables TX scheduling with
sub-ns accuracy (\ref{sec:crng-tosn:exp-tx-comp}). Therefore, the
response delay $\dtx$ and the additional delay $\delta_i$ for
responder identification in \crng can be enforced with high accuracy,
without relying on the timestamps embedded in the \response.

In more detail, the time of flight \tofi from the initiator to
responder \resp{i} is estimated as
\begin{equation}\label{eq:crng-tosn:tof}
  \tofi = \frac{\rtti - \dtxi}{2}
\end{equation}
and the corresponding distance as \mbox{$d_i = c \times \tofi$}. As
shown in Figure~\ref{fig:computedistance}, \dtxi is the delay between
the RX of \poll and the TX of \response at responder \resp{i}. This
delay is computed as the addition of three factors 
\mbox{$\dtxi = \dtx + \delta_i + \atx$}, where \dtx is
the fixed response delay inherited from \sstwr (\ref{sec:soa-sstwr}),
\mbox{$\delta_i = (i - 1)\tdelay$} is the responder-specific delay
enabling response identification (\ref{sec:crng-tosn:resp-id}), and
\atx is the known antenna delay obtained in a previous calibration 
step~\cite{dw1000-antenna-delay}.

\begin{figure}[!t]
\centering
\includegraphics[width=0.80\textwidth]{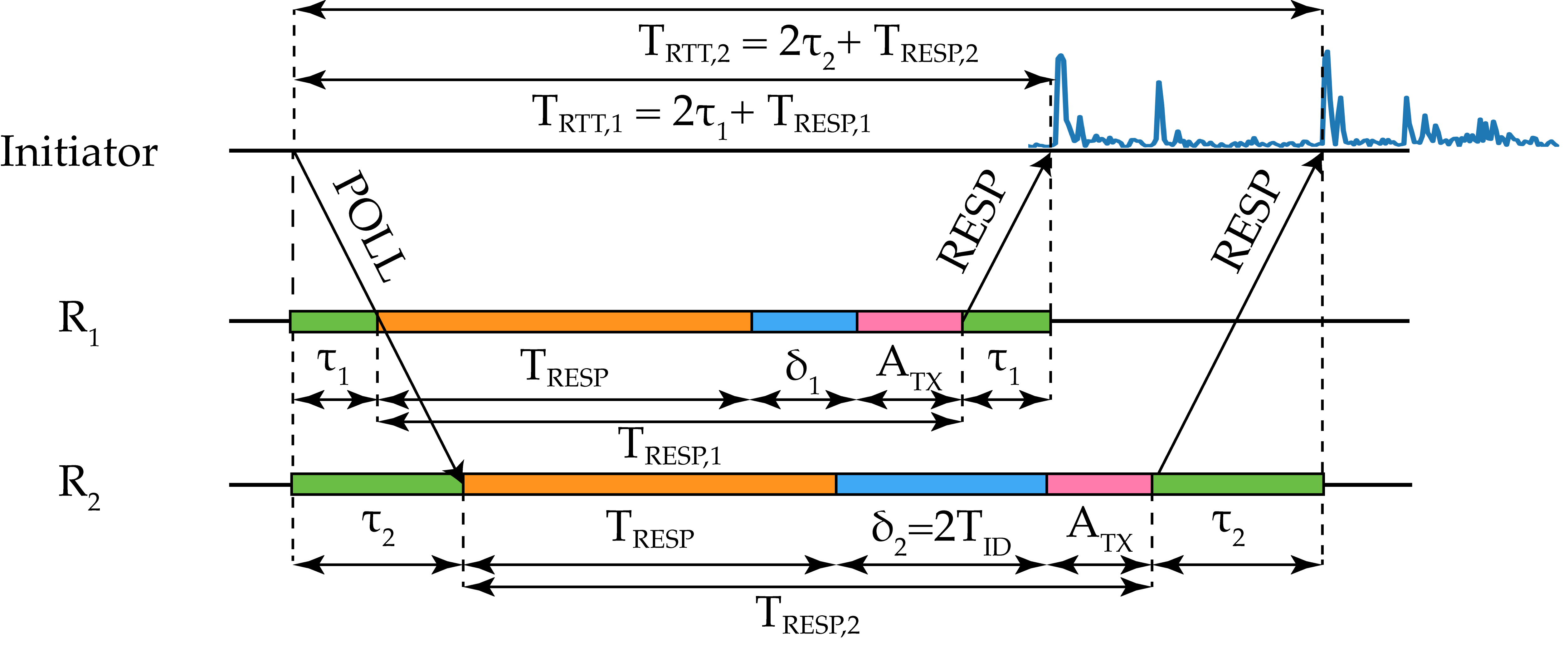}
\caption{Concurrent ranging time of flight $\tau_i$ computation. 
  To determine the distance
  $d_i = c \times \tau_i$ to responder \resp{i}, we need to accurately measure 
  the actual \response delay \mbox{$\dtxi = \dtx + \delta_i + \atx$} 
  and the round-trip time \rtti of each responder based on our \toa estimation.}
\label{fig:computedistance}
\end{figure}

\rtti is the round-trip time for responder \resp{i}, measured at the
initiator as the difference between the \response RX timestamp and the
\poll TX timestamp. The latter is accurately determined at the
\rmarker by the DW1000, in device time units of $\approx 15.65$~ps,
while the former must be extracted from the CIR via \toa
estimation. Nevertheless, the algorithms
in~\ref{sec:crng-tosn:toa-est} return only the CIR index \indexi at
which the first path of responder \resp{i} is estimated; this index
must therefore be translated into a radio timestamp, similar to the TX
\poll one.  To this end, we rely on the fact that the precise
timestamp \tfp associated to the \fpindex in the CIR is
known. Therefore, it serves as the accurate time baseline \wrt which to
derive the \response RX by
\begin{inparaenum}[\itshape i)] 
\item computing the difference $\Delta\indexfpi = \fpindex - \indexi$
  between the indexes in the CIR, and
\item obtaining the actual RX timestamp as
  $\tfp -  T_s\times\Delta\indexfpi$, where $T_s$ 
  is the CIR sampling period after upsampling (\ref{sec:crng-tosn:toa-est}).
\end{inparaenum}

In our experiments, we noticed that \crng usually underestimates
distance. This is due to the fact that the responder 
estimates the \toa of \poll with the DW1000
LDE algorithm, while the initiator estimates the \toa of each \response
with one of the algorithms in~\ref{sec:crng-tosn:toa-est}. For
instance, \ssub measures the \toa at the beginning of the path, while
LDE measures it at a peak height related to the noise standard
deviation reported by the DW1000.  This underestimation is nonetheless 
easily compensated by a constant offset ($\leq 20$~cm) whose value 
can be determined during calibration at deployment time.

Together, the steps we described enable accurate estimation of the
distance to multiple responders \emph{solely based on the CIR and the
  (single) RX timestamp provided by the radio}.  In the DW1000,
\begin{inparaenum}[\itshape i)]
\item the CIR is measured and available to the application even if RX
  errors occur, and 
\item the RX timestamp necessary to translate our \toa estimates to
  radio timestamps is always\footnote{Unless a very rare PHR error
    occurs~\cite[p.97]{dw1000-manual-v218}.} updated, 
\end{inparaenum}
therefore \emph{making our \crng prototype highly resilient to RX
  errors}.  Finally, the fact that we remove the dependency on
\resp{1} and therefore no longer need to embed/receive any timestamp
enables us to safely \emph{remove the entire payload from \response
  packets}. Unless application information is piggybacked on a
\response, this can be composed only of preamble, SFD, and PHR,
reducing the length of the \response packet, and therefore 
the latency and energy consumption of \crng.

\section{Evaluation}
\label{sec:crng-tosn:eval}

We evaluate our \crng prototype, embodying the techniques illustrated
in~\ref{sec:reloaded}. We begin by describing our experimental setup
(\ref{sec:crng-exp-setup}) and evaluation metrics
(\ref{sec:exp-metrics}). Then, we evaluate our
TX scheduling (\ref{sec:crng-tosn:exp-tx-comp}), confirming its
ability to achieve sub-ns precision. This is key to improve the
accuracy of ranging and localization, which we evaluate in
static positions (\ref{sec:crng-exp-static}) and via trajectories
generated by a mobile robot in an OptiTrack facility
(\ref{sec:optitrack}).

\subsection{Experimental Setup}\label{sec:crng-exp-setup}
We implemented \crng atop Contiki OS~\cite{contikiuwb} using the 
EVB1000 platform~\cite{evb1000} as in~\ref{sec:questions}.

\fakeparagraph{UWB Radio Configuration} In all experiments, we set the
DW1000 to use channel~7 with center frequency $f_c = 6489.6$~GHz and
$900$~MHz receiver bandwidth. We use the shortest preamble length of
64~symbols with preamble code~17, the highest $PRF = 64~$MHz, and the
highest 6.8~Mbps data rate. Finally, we set the response delay
$\dtx = $~\SI{800}{\micro\second} to provide enough time to compensate
for the TX scheduling uncertainty (\ref{sec:crng-tosn:txfix}).

\fakeparagraph{Concurrent Ranging Configuration}
Table~\ref{tab:crng-tosn:parameters} summarizes the default values of
\crng parameters. The time shift $\tdelay = 128$~ns for \response
identification (\ref{sec:crng-tosn:resp-id}) corresponds to a distance
of $38.36$~m, sufficiently larger than the maximum distance difference
($\approx 12$~m) among anchors in our setups.  For \toa estimation
(\ref{sec:crng-tosn:toa-est}), we use a noise threshold 
$\threshold = 11 \times \stdnoise$, computed as described
in~\ref{sec:crng-tosn:cir-proc}, and $\niterations = 3$ iterations per
CIR chunk of the \ssub algorithm.

\begin{table}[!t]
\centering
  \caption{Main parameters of concurrent ranging with default values.}
  \label{tab:crng-tosn:parameters}
\begin{tabular}{llr}
  \toprule
  {\bfseries Symbol} & {\bfseries Description} & {\bfseries Default Value}\\
  \midrule
  $L$ & CIR upsampling factor & 30\\
  $\tdelay$ & Time shift for response identification & 128~ns\\
  $\xi$ & Noise threshold for CIR re-arrangement & 0.14\\
  $W$ & Window length for CIR re-arrangement & 228~samples\\
  $\threshold$ & Noise threshold for \toa estimation algorithm & $11\times \stdnoise$\\
  $\niterations$ & Iterations (max. number of paths) of the \ssub \toa algorithm & 3\\
  \bottomrule
\end{tabular}
\end{table}

\fakeparagraph{Infrastructure} 
We run our experiments with a mobile testbed infrastructure 
we deploy in the target environment. 
Each testbed node consists of an EVB1000~\cite{evb1000} connected via USB to a
Raspberry~Pi (RPi)~v3, equipped with an ST-Link programmer enabling
firmware uploading. Each RPi reports its serial data via WiFi to a
server, which stores it in a log file. 
Although our prototype supports runtime positioning, 
hereafter we run our analysis offline.

In each test, we collect TX information from anchors and RX
information diagnostics and CIR signals from the initiator.  We
collect a maximum of 8~CIR signals per second, as this requires
reading over SPI, logging over USB, and transmitting over WiFi the
4096B accumulator buffer (CIR) together with the rest of the measurements.

\fakeparagraph{Baseline: SS-TWR with and without Clock Drift Compensation} 
We compare the performance of \crng against the
commonly-used SS-TWR scheme (\ref{sec:soa-sstwr}).  We implemented
it for the EVB1000 platform atop Contiki OS using a response delay
$\dtx = \SI{320}{\micro\second}$ to minimize the impact of clock
drift. Moreover, we added the possibility to compensate for the
estimated clock drift at the initiator based on the carrier frequency
offset (CFO) measured during the \response packet RX as suggested by
\decawave~\cite{dw-cfo, dw1000-sw-api}. Hence, our evaluation results also
serve to quantitatively demonstrate the benefits brought by this
recent clock drift compensation mechanism. 
As for localization, we perform a \sstwr exchange every 2.5~ms against
the $N$ responders deployed, in round-robin, yielding an estimate of
the initiator position every $N \times 2.5$~ms. We use the exact same
RF configuration as in \crng, for comparison.

\subsection{Metrics}
\label{sec:exp-metrics}
Our main focus is on assessing the ranging and localization accuracy
of \crng in comparison with \sstwr. Therefore, we consider the
following metrics, for which we report the median, average $\mu$, and
standard deviation $\sigma$, along with various percentiles of the
absolute values:

\begin{itemize}
\item \emph{Ranging Error.} We compute it \wrt each responder $R_i$ as
  $\hat{d}_{i} - d_{i}$, where $\hat{d}_{i}$ is the distance estimated 
  and $d_{i}$ is the known distance. 

\item \emph{Localization Error.} We compute the absolute positioning
  error as $\norm{\mathbf{\hat p} - \mathbf{p_r}}$, where
  $\mathbf{\hat p}$ is the initiator position estimate and
  $\mathbf{p_r}$ its known position.
\end{itemize}

Moreover, we also consider the \emph{success rate}
as a measure of the reliability and robustness of \crng in real
environments. Specifically, we define the \textit{ranging success
  rate} to responder \resp{i} and the \textit{localization success
  rate} as the fraction of CIR signals where, respectively, we are able to
\begin{inparaenum}[\itshape i)]
\item measure the distance $d_i$ from the initiator to \resp{i} and
\item obtain enough information \mbox{($\geq 3$ \toa estimates)} to
  compute the initiator position $\mathbf{\hat p}$.
\end{inparaenum}

  \subsection{Precision of TX Scheduling}
\label{sec:crng-tosn:exp-tx-comp}

\begin{figure}[!t]
\centering  
\includegraphics{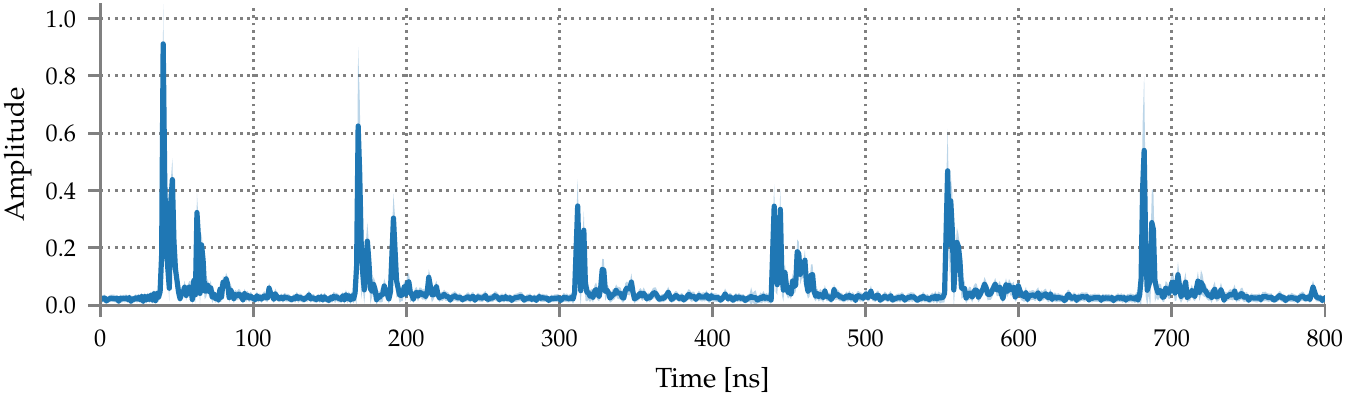}   
\caption{Average CIR amplitude and standard deviation per time delay
across 500 signals with the initiator in the left center position of
Figure~\ref{fig:crng-tosn:err-ellipse-disi}.}
\label{fig:crng-cir-meanstd}
\end{figure}

We begin by examining the ability of our TX compensation mechanism
(\ref{sec:crng-tosn:txfix}) to schedule transmissions
precisely, as this is crucial to improve the
accuracy of concurrent ranging and localization.  To this end, we ran an
experiment with one initiator and six responders, collecting 500 CIR
signals for our analysis.  Figure~\ref{fig:crng-cir-meanstd} shows the
average CIR amplitude and standard deviation after re-arranging the
CIRs (\ref{sec:crng-tosn:cir-rearrangement}) and aligning the
upsampled CIR signals based on the direct path of responder \resp{1}. 
Across all time delays, the average CIR presents only minor amplitude
variations in the direct paths and MPC.  Further, the precise scheduling of
\response transmissions yields a high amplitude for the direct paths
of all signals; this is in contrast with the smoother and flatter
peaks we observed earlier (\ref{sec:questions},
Figure~\ref{fig:single-tx-cir-variations}) due to the TX uncertainty
$\epsilon \in [-8, 0)$~ns.

To quantitatively analyze the TX precision, we estimate the \toa of
each \response and measure the time difference $\Delta t_{j, 1}$
between the \toa of responder \resp{j} and the one of \resp{1}, chosen
as reference, after removing the time delays $\delta_i$ used 
for response identification. Then, we subtract the mean of the distribution
and look at the deviations of $\Delta t_{j, 1}$, which ideally 
should be negligible.
Figure~\ref{fig:crng-ts-cdf} shows the CDF of the $\Delta t_{j, 1}$
variations from the mean, while Table~\ref{tab:crng-tx-dev} details
the percentiles of the absolute variations. All time differences
present a similar behavior with an aggregate mean error
$\mu = 0.004$~ns across the 2,500 $\Delta t_{j, 1}$ measurements, with
$\sigma = 0.38$~ns and a median of $0.03$~ns; the absolute \nth{90},
\nth{95}, and \nth{99} percentiles are 0.64, 0.77, and 1.09~ns,
respectively. These results confirm that our implementation is able
to \emph{reliably schedule transmissions with sub-ns precision}.

\begin{figure}[!t]
\centering
\includegraphics{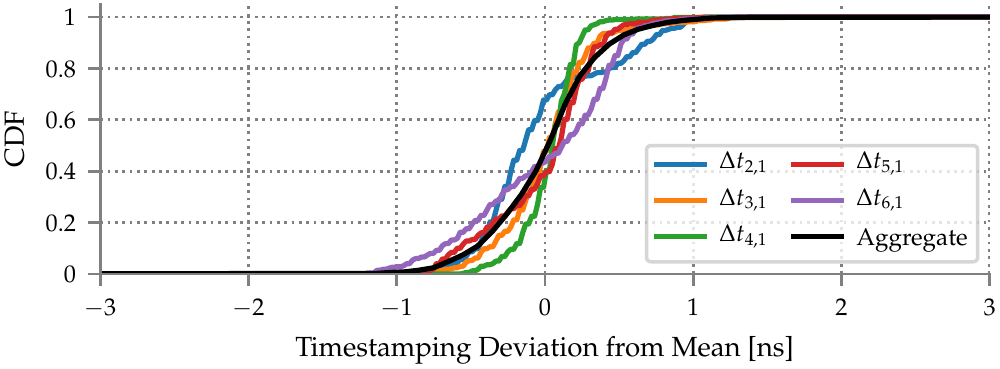}   
\caption{Time difference deviation from the mean across 500 CIRs.}
\label{fig:crng-ts-cdf}
\end{figure}

\begin{table}[h]
\centering
  \caption{Deviation percentiles for the absolute time difference 
  $\Delta t_{j,1}$ variations.}
  \label{tab:crng-tx-dev}
\begin{tabular}{ccccccc}
\toprule
& \multicolumn{6}{c}{\bfseries Percentile [ns]}\\
\cmidrule(lr){2-7}
{\bfseries Time Difference} & {\nth{25}} & {\nth{50}} & {\nth{75}} & {\nth{90}} 
& {\nth{95}} & {\nth{99}}\\
\midrule
$\Delta t_{2,1}$ & 0.15 & 0.32 & 0.52 & 0.72 &0.85 & 1.08\\
$\Delta t_{3,1}$ & 0.08 & 0.18 & 0.32 & 0.51 &0.68 & 1.12\\
$\Delta t_{4,1}$ & 0.06 & 0.13 & 0.20 & 0.30 &0.40 & 0.60\\
$\Delta t_{5,1}$ & 0.13 & 0.23 & 0.40 & 0.64 &0.74 & 0.90\\
$\Delta t_{6,1}$ & 0.24 & 0.39 & 0.54 & 0.76 &0.91 & 1.14\\
Aggregate & 0.10 & 0.23 & 0.42 & 0.64 &0.77 & 1.09\\
\bottomrule
\end{tabular}
\end{table}

  \subsection{Performance with Static Targets}
\label{sec:crng-exp-static}

We report the results from experiments in a $6.4 \times 6.4$~m$^2$
area inside our office building, using 6~concurrent responders that 
serve as localization anchors. We place the initiator in
18~different positions and collect 500~CIR signals at each of them,
amounting to 9,000~signals.

The choice of initiator positions is key to our analysis. As shown in
Figure~\ref{fig:crng-tosn:err-ellipse-disi}, we split the 18~positions
in two halves with different purposes. The 9~positions in the center
dashed square are representative of the positions of interest for most
applications, as they are farther from walls and enjoy the best
coverage \wrt responders, when these serve as anchors.  Dually, the
remaining 9~positions can be regarded as a stress test of sorts. They
are very close to walls, yielding significant MPC; this is an issue
with conventional \sstwr but is exacerbated in \crng, as it increases
the possibility to confuse MPC with the direct paths of
responders. Further, these positions are at the edge of the area
delimited by anchors, therefore yielding a more challenging geometry
for localization. 
Hereafter, we refer to these two sets of
positions as \centerpos and \edgepos, respectively, and analyze the
performance in the common case represented by \centerpos as well as in
the more challenging, and somewhat less realistic, case where
\emph{all} positions are considered.

In each position, we measure the ranging and localization performance
of \crng with both our \toa estimation algorithms
(\ref{sec:crng-tosn:toa-est}) and compare it, in the same setup,
against the performance of the two \sstwr variants we consider.

\fakeparagraph{Ranging Accuracy}
Figure~\ref{fig:crng-tosn:center-rng-err-cdf} shows the CDF of the
ranging error \mbox{$\hat{d}_i - d_i$} obtained with \crng and \sstwr
in \centerpos positions; Table~\ref{tab:rng-err-center} offers an
alternate view by reporting the values of the metrics we consider
(\ref{sec:exp-metrics}).

The performance of \crng in this setting, arguably the one of interest
for most applications, is remarkable and in line with the one of
\sstwr. All variants achieve a similar centimeter-level median and
average error. Although \sstwr exhibits a smaller $\sigma$, both \crng
and \sstwr achieve decimeter-level precision. This is also reflected
in the absolute error, which is nonetheless very small. Both variants
of \crng achieve $\nth{99} = 28$~cm, only a few cm higher than plain
\sstwr, while its drift compensated variant achieves a lower
$\nth{99} = 18$~cm. The latter \sstwr variant is the technique that,
as expected, achieves the best results across the board. 
Nevertheless, \crng measures the distance to the $N=6$ 
responders concurrently, reducing the number of two-way exchanges 
from~6 to~1, therefore providing a significant reduction 
in channel utilization and other evident benefits in terms 
of latency, energy, and scalability. 
Interestingly, the difference in accuracy and precision
between the two \crng variants considered is essentially negligible.

\begin{figure}[!t]
\centering
  \subfloat[\centerpos positions.\label{fig:crng-tosn:center-rng-err-cdf}]{
 \includegraphics{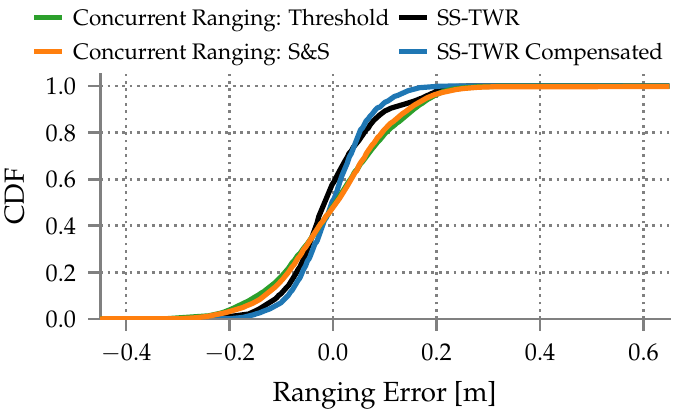}   
}
\subfloat[All positions.\label{fig:crng-tosn:all-rng-err-cdf}]{
  \includegraphics{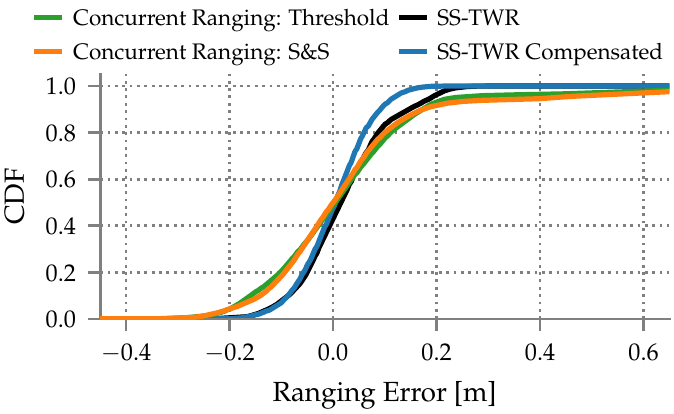}   
}
\caption{CDF of ranging error with static positions.}
\label{fig:crng-tosn:center-err-hist-disi}
\end{figure}

\begin{table}
\centering
  \caption{Ranging error comparison across the 9 \centerpos positions
  considered.}
  \label{tab:rng-err-center}
\begin{tabular}{l ccc ccccc}
\toprule
& \multicolumn{3}{c}{ $\hat{d}_i - d_i$ [cm]}
  & \multicolumn{5}{c}{ $|\hat{d}_i - d_i|$ [cm]}\\
\cmidrule(lr){2-4} \cmidrule(lr){5-9}
{\bfseries Scheme} & Median & {$\mu$} & {$\sigma$}
  & \nth{50} & \nth{75} & \nth{90} & \nth{95} & \nth{99}\\
\midrule
Concurrent Ranging: Threshold & 0.4 & 0.3 & 11.9 & 8 &14 & 19 &21 & 28\\
Concurrent Ranging: \ssub & 0.7 & 0.5 & 11.7 & 7 &12 & 18 &21 & 28\\
\sstwr & -1.7 & -0.5 & 8.6  &5 &9 & 15 &19 & 22\\
\sstwr Compensated & -0.3 & -0.3 & 6.9 &4 &8 & 12 &14 & 18\\
\bottomrule
\end{tabular}
\end{table}

\begin{table}
\centering
\caption{Ranging error comparison across the 18 static positions
  considered (both \centerpos and \edgepos).}
  \label{tab:rng-err-all}
\begin{tabular}{l ccc ccccc}
\toprule
& \multicolumn{3}{c}{ $\hat{d}_i - d_i$ [cm]}
  & \multicolumn{5}{c}{ $|\hat{d}_i - d_i|$ [cm]}\\
\cmidrule(lr){2-4} \cmidrule(lr){5-9}
{\bfseries Scheme} & Median & {$\mu$} & {$\sigma$}
  & \nth{50} & \nth{75} & \nth{90} & \nth{95} & \nth{99}\\
\midrule
Concurrent Ranging: Threshold & 0.4 & 2.0 & 17.7 &9 &15 & 21 &28 & 81\\
Concurrent Ranging: \ssub & 0.1 & 3.1 & 20.4 &8 &14 & 23 &44 & 91\\
\sstwr & 1.5 & 2.1 & 8.8 &6 &10 & 16 &19 & 23\\
\sstwr Compensated & 0.4 & 0.2 & 6.9 &5 &8 & 12 &14 & 18\\
\bottomrule
\end{tabular}
\end{table}

Figure~\ref{fig:crng-tosn:all-rng-err-cdf} shows instead the CDF of
the ranging error across all positions, \ie both \centerpos and
\edgepos, while Table~\ref{tab:rng-err-all} shows the values of the
metrics we consider. The difference in accuracy between the two \crng
variants is still negligible in aggregate terms, but slightly worse
for \ssub when considering the absolute error; this is balanced by a
higher reliability \wrt the threshold-based variant, as discussed
later. In general, the accuracy of \crng is still comparable to the
\centerpos case in terms of median and average error, although with
slightly worse precision. This is also reflected in the absolute
error, which remains very small and essentially the same as in the
\centerpos case until the $75^\mathit{th}$ percentile, but reaches
$\nth{99} = 91$~cm with \ssub. In contrast, the performance of both
variants of \sstwr is basically unaltered.

These trends can also be observed in the alternate view of
Figure~\ref{fig:crng-tosn:err-hist-disi}, based on normalized
histograms. The distributions of \crng and \sstwr are similar,
although the latter is slightly narrower. Nevertheless, \crng has a
small tail of positive errors, not present in \sstwr, yielding higher
values of $\sigma$ and $\geq\nth{90}$ percentiles in
Table~\ref{tab:rng-err-all}. Further, these tails are also not present
in the case of \centerpos, whose distribution is otherwise essentially
the same, and therefore not shown due to space limitations.

\begin{figure}[!t]
\centering
\subfloat[Threshold-based \toa estimation.\label{fig:crng-tosn:err-hist-th}]{
 \includegraphics{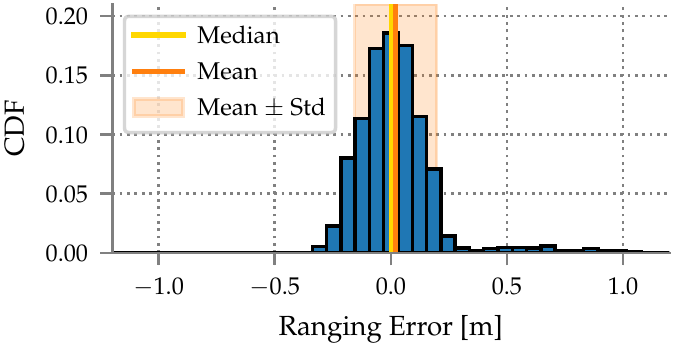}   
}
\subfloat[\ssub with $K = 3$ iterations.\label{fig:crng-tosn:err-hist-ssr3}]{
  \includegraphics{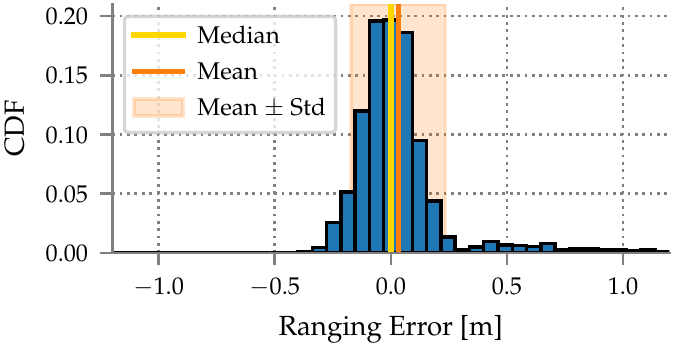}   
}\\
\subfloat[\sstwr.\label{fig:crng-tosn:err-hist-sstwr}]{
 \includegraphics{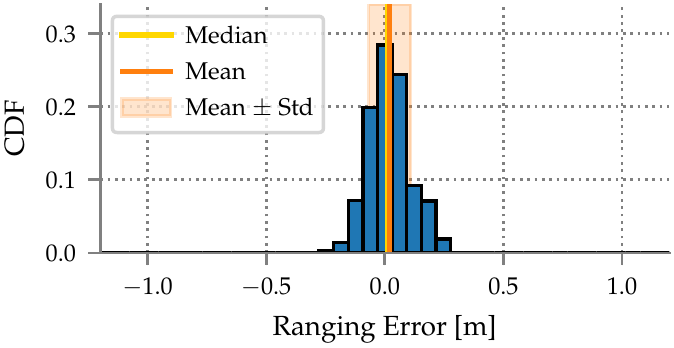}   
}
\subfloat[\sstwr with drift compensation.\label{fig:crng-tosn:err-hist-sstwr-drift}]{
  \includegraphics{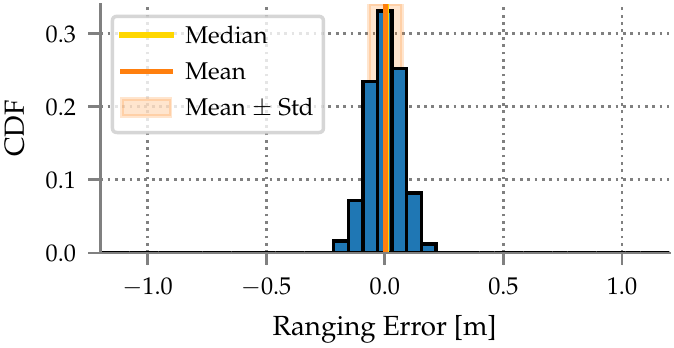}
}
\caption{Normalized histogram of ranging error across all 18 static
  positions (both \centerpos and \edgepos).}
\label{fig:crng-tosn:err-hist-disi}
\end{figure}

This is to be ascribed to \edgepos positions, in which the initiator
\begin{inparaenum}[\itshape i)]
\item is next to a wall suffering from closely-spaced and strong MPC 
next to the direct path, and
\item is very close to one or two anchors and far from the others, 
resulting in significantly different power loss across responses. 
\end{inparaenum}
This setup sometimes causes the direct path of some responses to be
buried in MPC noise or even unable to cross the noise threshold
$\threshold$.  As a result, our \toa algorithms erroneously select one
of the MPC peaks as the first path, yielding an incorrect distance
estimate. Nevertheless, as mentioned, the absolute error remains
definitely acceptable with both the threshold-based and \ssub \toa
algorithms. 

\fakeparagraph{Localization Accuracy}
Figure~\ref{fig:crng-tosn:err-ellipse-disi} shows the localization
error and $3\sigma$ ellipses for each initiator position and both \toa
estimation algorithms, while
Table~\ref{tab:loc-err-center}--\ref{tab:loc-err-all} show the values
of the metrics we consider. 
Coherently with the analysis of ranging accuracy, the standard
deviation $\sigma$ for \crng is significantly lower in the \centerpos
positions than in the \edgepos ones. This is a consequence of the
distance overestimation we observed, which causes larger ellipses and
a small bias \wrt the true position in a few \edgepos
positions. Interestingly, both \toa algorithms underperform in the
same positions, although sometimes with different effects, \eg in
positions $(1.6,-3.2)$ and $(3.2,-1.6)$.

The difference between \sstwr and \crng is also visible in the longer
tails of the localization error CDF (Figure~\ref{fig:crng-tosn:cdf-disi-all}), 
where it is further exacerbated by the fact that, in our setup, 
the worst-case \edgepos positions are \emph{as many as} 
the common-case \centerpos ones.
Nevertheless, even in this challenging case,
Table~\ref{tab:loc-err-all} shows that \crng still achieves
decimeter-level accuracy, with the median\footnote{As the localization
  error is always positive, unlike the ranging error, the median is
  the same as the \nth{50} percentile.}  nearly the same as plain
\sstwr. The error is also quite small; $\nth{75}\leq 17$~cm
and $\nth{99} \leq 57$~cm, with the threshold-based approach
performing marginally better than \ssub, as in ranging. However, the
drift compensated \sstwr is still the most accurate and precise.

\begin{figure}[!t]
\centering
\subfloat[Threshold-based \toa estimation.\label{fig:crng-err-ellipse-th}]{
 \includegraphics{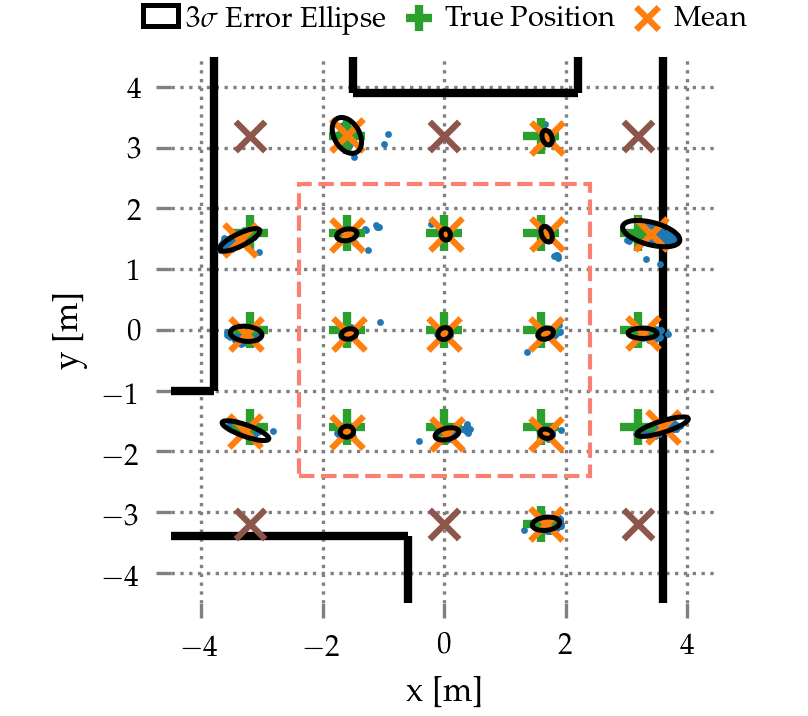}   
}
\subfloat[\ssub with $K = 3$ iterations.\label{fig:crng-err-ellipse-ssr}]{
  \includegraphics{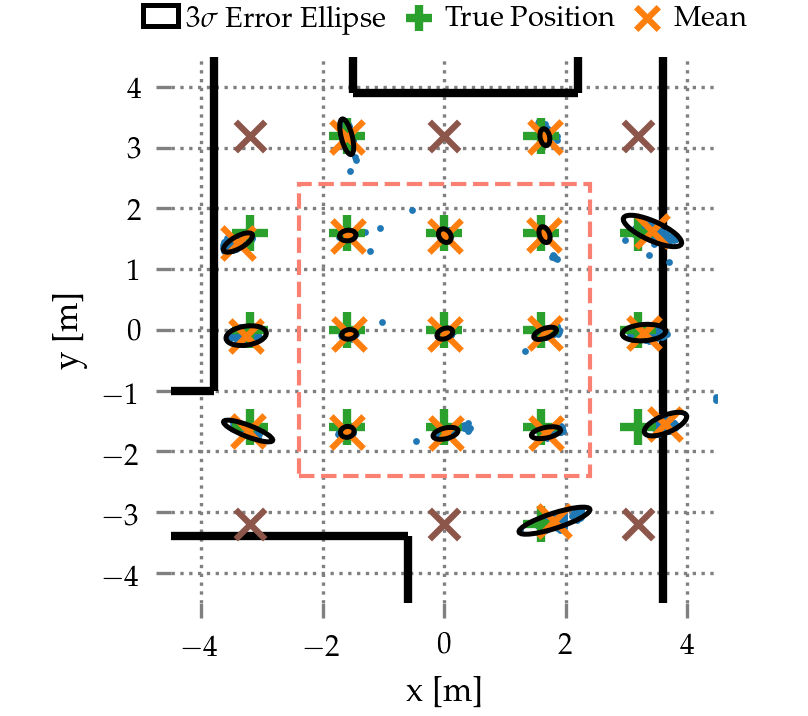}   
}
\caption{$3\sigma$ error ellipses with \crng and six concurrent responders.
Blue dots represent position estimates, brown crosses are anchors.
The dashed light red square denotes the positions of interest.}
\label{fig:crng-tosn:err-ellipse-disi}
\end{figure}

\begin{figure}[!t]
\centering
\subfloat[\centerpos positions.\label{fig:crng-tosn:cdf-disi-center}]{
  \includegraphics{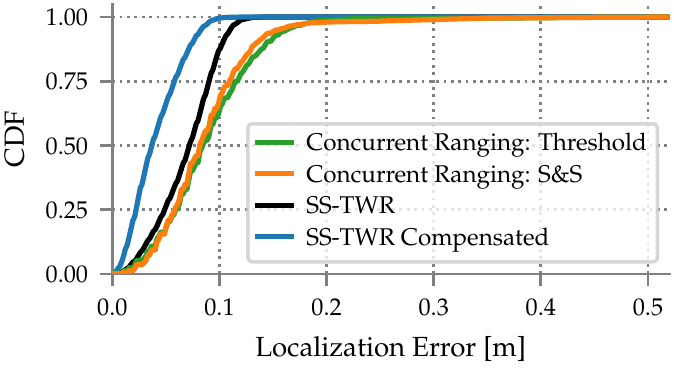}
}
\subfloat[All positions.\label{fig:crng-tosn:cdf-disi-all}]{
 \includegraphics{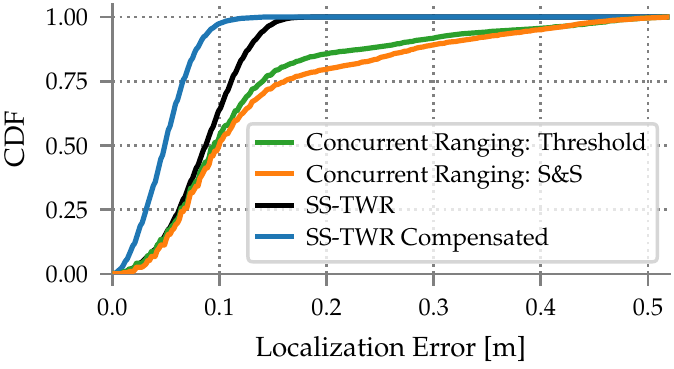}
}
\caption{CDF of localization error in static positions.}
\label{fig:crng-tosn:cdf-disi}
\end{figure}

The gap with \sstwr further reduces in the more common
\centerpos positions, where the accuracy of \crng is very high,
as shown in Figure~\ref{fig:crng-tosn:err-ellipse-disi} and
Figure~\ref{fig:crng-tosn:cdf-disi-center}. Position estimates are
also quite precise, with $\sigma \leq 5$~cm. Further, the error
remains $\leq 16$~cm in $95\%$ of the cases, regardless of the \toa
estimation technique; the threshold-based and \ssub \toa algorithms
show only a marginal difference, with a \nth{99} percentile of $21$~cm
and $30$~cm, respectively.

\begin{table}
\centering
  \caption{Localization error comparison across the 9 \centerpos positions
  considered.}
  \label{tab:loc-err-center}
\begin{tabular}{l cc ccccc}
\toprule
& \multicolumn{7}{c}{$\norm{\mathbf{\hat p} - \mathbf{p_r}}$ [cm]}\\
\cmidrule(lr){2-8}
{\bfseries Scheme} & {$\mu$} & {$\sigma$} 
  & \nth{50} & \nth{75} & \nth{90} & \nth{95} & \nth{99}\\
\midrule
Concurrent Ranging: Threshold & 9 & 4.9 &8 &12 & 14 &16 & 21\\
Concurrent Ranging: \ssub & 8.8 & 5 &8 &11 & 14 &16 & 30\\
\sstwr & 6.9 & 2.7 &7 &9 & 10 &11 & 12\\
\sstwr Compensated & 4.1 & 2.3 &4 &6 & 8 &8 & 10\\
\bottomrule
\end{tabular}
\end{table}

\begin{table}
\centering
  \caption{Localization error comparison across the 18 static positions
  considered (both \centerpos and \edgepos).}
  \label{tab:loc-err-all}
\begin{tabular}{l cc ccccc}
\toprule
& \multicolumn{7}{c}{$\norm{\mathbf{\hat p} - \mathbf{p_r}}$ [cm]}\\
\cmidrule(lr){2-8}
{\bfseries Scheme} & {$\mu$} & {$\sigma$} 
  & \nth{50} & \nth{75} & \nth{90} & \nth{95} & \nth{99}\\
\midrule
Concurrent Ranging: Threshold & 12.9 & 11 &10 &14 & 28 &41 & 51\\
Concurrent Ranging: \ssub & 14.5 & 12.6 &10 &17 & 33 &42 & 57\\
\sstwr & 8.6 & 3.4 &9 &11 & 13 &14 & 16\\
\sstwr Compensated & 5 & 2.4 &5 &7 & 8 &9 & 11\\
\bottomrule
\end{tabular}
\end{table}

\fakeparagraph{Success Rate} Across the 9,000 CIR signals considered
in this section, \crng is able to extract a position estimate in 8,663
and 8,973 of them using our threshold-based and \ssub \toa estimation,
respectively, yielding a remarkable localization success rate of
$96.25\%$ and $99.7\%$. Across the successful estimates, 6~samples
included very large errors $\geq 10$~m. These could be easily
discarded with common filtering techniques~\cite{ukf-julier}.  In the
\centerpos positions of interest, the localization success rate with
both \toa techniques yields $99.7\%$. 

Threshold-based \toa estimation is more susceptible to strong and late
MPC occurring at the beginning of the following CIR chunk, which
result in invalid distance estimates that are therefore discarded,
reducing the success rate. 
As for \ssub, of the $27$ signals failing to provide an estimate, $21$
are caused by PHR errors where the DW1000 does not update the RX
timestamp. In the remaining 6~signals, \ssub was unable to detect the
first or last responder; these signals were therefore discarded, to
avoid a potential responder mis-identification
(\ref{sec:crng-tosn:toa-est}).

Regarding ranging, threshold-based estimation yields a success rate of
$95.98\%$ across the 54,000 expected estimates, while \ssub reaches
$99.58\%$, in line with the localization success rate.

  \subsection{Performance with Mobile Targets}
\label{sec:optitrack}

We now evaluate the ability of \crng to accurately determine the
position of a mobile node. This scenario is representative of several
real-world applications, \eg exploration in harsh
environments~\cite{thales-demo}, drone operation~\cite{guo2016ultra},
and user navigation in museums or shopping
centers~\cite{museum-tracking}.

To this end, we ran experiments with an EVB1000 mounted on a mobile
robot~\cite{diddyborg} in a $12 \times 8$~m$^2$ indoor area where we
placed 6~responders serving as localization anchors. 
We compare both our \crng variants against only \sstwr with clock drift 
compensation, as this provides a more challenging baseline, 
as discussed earlier.
The area is equipped with 14~OptiTrack cameras~\cite{optitrack}, which we
configured to output positioning samples with an update rate of
125~Hz and calibrated to obtain a mean 3D error $< 1$~mm, 
therefore yielding reliable and accurate ground truth to
validate the UWB systems against. The mobile robot is controlled by a
RPi, enabling us to easily repeat trajectories by remotely driving the
robot over WiFi via a Web application on a smartphone. A second RPi enables the
flashing of the EVB1000 node with the desired binary and the upload of
serial output (CIRs and RX information) to our testbed server for
offline analysis.

\begin{figure}[!t]
\centering
 \includegraphics{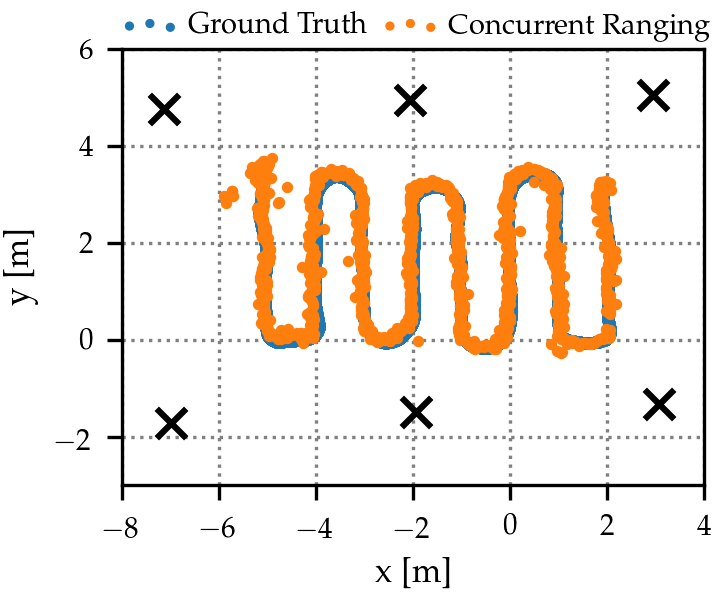}   
\hspace{5mm}
 \includegraphics{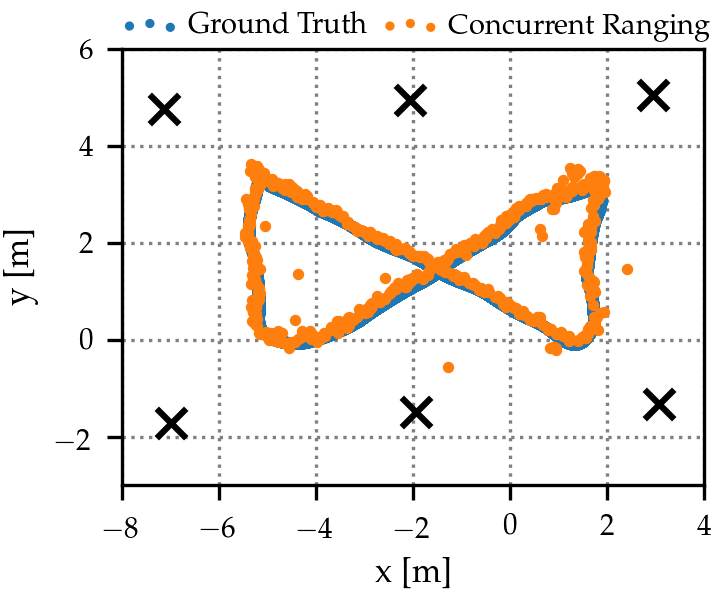}   
\\
 \includegraphics{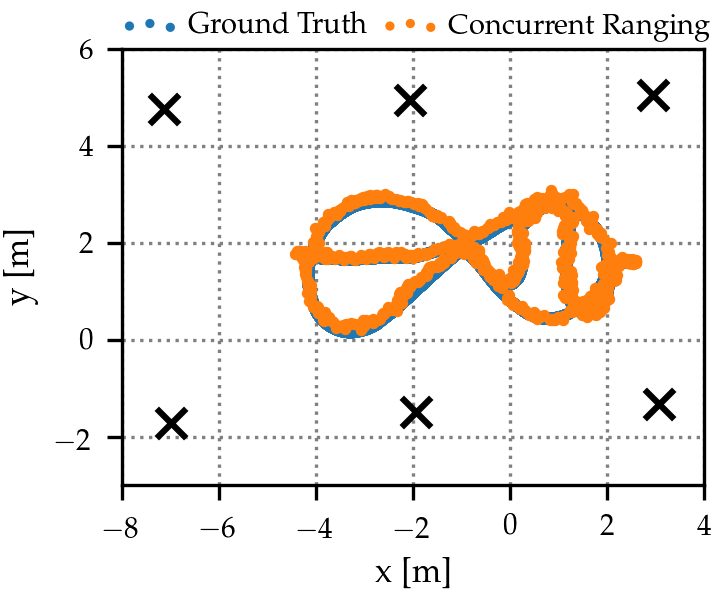}   
\hspace{5mm}
 \includegraphics{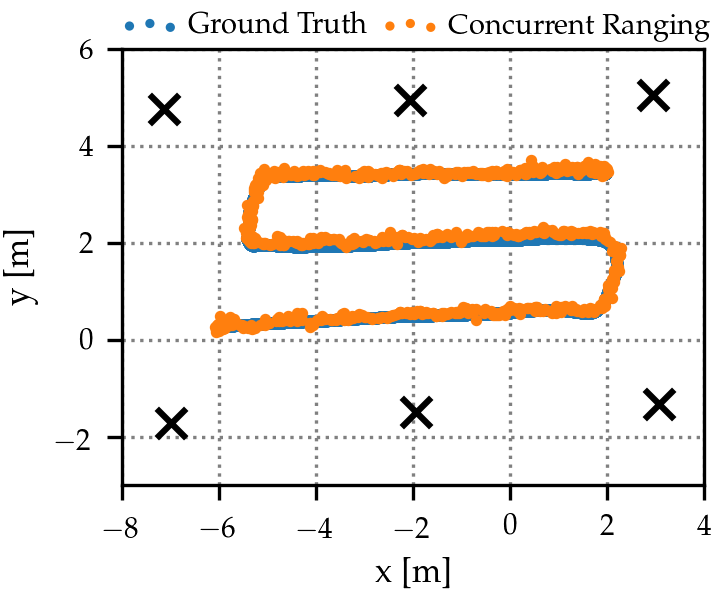}   
\caption{Localization with \crng across four trajectories using \ssub 
  with $K = 3$ iterations.}
\label{fig:crng-tosn:tracking}
\end{figure}

Before presenting in detail our evaluation,
Figure~\ref{fig:crng-tosn:tracking} offers the opportunity to visually
ascertain that our \crng prototype is able to \emph{continuously and
  accurately} track the robot trajectory, by comparing it against the
ground truth obtained with OptiTrack. We observe a few position
samples with relatively high error, due to strong MPC; however, these
situations are rare and, in practice, easily handled with techniques
commonly used in position tracking, \eg extended or unscented Kalman
filters~\cite{ukf-julier}. Due to space constraints, the figure shows
only trajectories with \ssub because they are very similar to
threshold-based ones, as discussed next.

\fakeparagraph{Ranging Accuracy} 
Across all samples, we compute the ranging error 
$\hat{d}_{i} - d_{i}$ between the \crng or \sstwr
estimate $\hat{d}_{i}$ for \resp{i} and the OptiTrack estimate $d_{i}$. 
To obtain the latter, we interpolate the high-rate positioning traces
of OptiTrack to compute the exact robot position $\mathbf{p}$ at each 
time instance of our \crng and \sstwr traces and then estimate
the true distance $d_i = \norm{\mathbf{p} - \mathbf{p_i}}$, where
$\mathbf{p_i}$ is the known position of \resp{i}.

Table~\ref{tab:optitrack-rng-err} shows that the results exhibit the
very good trends we observed in the static case
(\ref{sec:crng-exp-static}).  In terms of accuracy, the median and
average error are very small, and very close to \sstwr. 
However, \sstwr is significantly more precise, while the
standard deviation $\sigma$ of \crng is in line with the one observed
with all 18~positions (Table~\ref{tab:rng-err-all}). 
In contrast, however, the absolute error is
$\nth{99}\leq 37$~cm, significantly lower than in this latter
case. Further, the \toa algorithm employed for \crng has only a
marginal impact on accuracy and precision.

\begin{figure}[!t]
\centering
\subfloat[Concurrent ranging: Threshold.\label{fig:crng-tosn:pergine-th-err-hist}]{
 \includegraphics{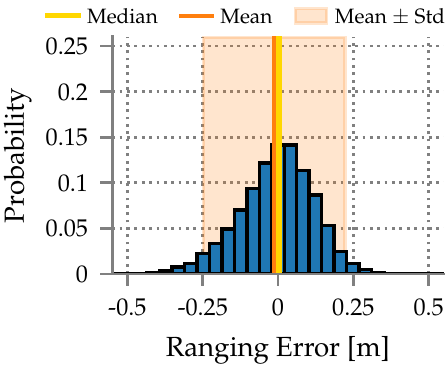}   
}
\subfloat[Concurrent ranging: \ssub.\label{fig:crng-tosn:pergine-ss3-err-hist}]{
  \includegraphics{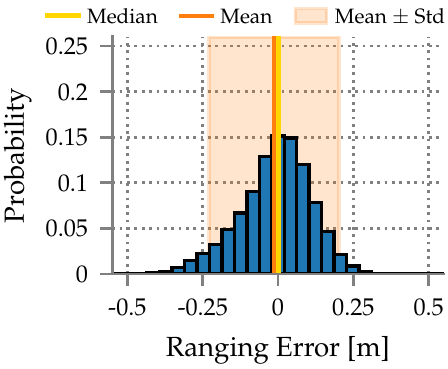}   
}
\subfloat[\sstwr with drift compensation.\label{fig:crng-tosn:pergine-sstwr-err-hist}]{
 \includegraphics{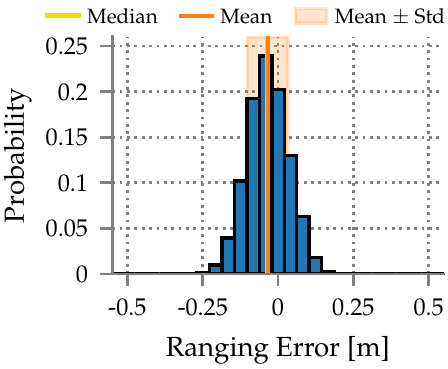}   
}
\caption{Normalized histogram of the ranging error across multiple
  mobile trajectories.}
\label{fig:crng-tosn:pergine-rng-err-hist}
\end{figure}

\begin{table}[!t]
\centering
  \caption{Ranging error comparison across multiple mobile trajectories.}
  \label{tab:optitrack-rng-err}
  \begin{tabular}{l ccc ccccc}
\toprule
& \multicolumn{3}{c}{ $\hat{d}_i - d_i$ [cm]}
  & \multicolumn{5}{c}{ $|\hat{d}_i - d_i|$ [cm]}\\
\cmidrule(lr){2-4} \cmidrule(lr){5-9}
{\bfseries Scheme} & Median & {$\mu$} & {$\sigma$} 
  & \nth{50} & \nth{75} & \nth{90} & \nth{95} & \nth{99}\\
\midrule
Concurrent Ranging: Threshold & 0.3 & -1.3 & 23.5 &8 &14 & 20 &25 & 37\\
Concurrent Ranging: \ssub & 0.2 & -1.4 & 21.6 &8 &13 & 20 &24 & 35\\
\sstwr Compensated & -3.5 & -3.4 & 6.8 &5 &9 & 12 &15 & 19\\
\bottomrule
\end{tabular}
\end{table}

An alternate view confirming these observations is
offered by the normalized histograms in
Figure~\ref{fig:crng-tosn:pergine-rng-err-hist}, where
the long error tails observed in 
Figure~\ref{fig:crng-tosn:err-hist-th}--\ref{fig:crng-tosn:err-hist-ssr3}
are absent in
Figure~\ref{fig:crng-tosn:pergine-th-err-hist}--\ref{fig:crng-tosn:pergine-ss3-err-hist}.

Overall, \crng follows closely the performance of \sstwr with drift
compensation, providing a more scalable scheme with less overhead
and comparable accuracy. Notably, \crng measures the
distance to all responders simultaneously, an important factor
when tracking rapidly-moving targets to reduce the bias induced by
relative movements. Further, this aspect also enables a
significant increase of the attainable update rate.

\fakeparagraph{Localization Accuracy}
Figure~\ref{fig:crng-optitrack-cdf} compares the 
CDFs of the localization error of the techniques under evaluation;
Table~\ref{tab:optitrack-loc-err} reports the value of the metrics
considered. The accuracy of \sstwr is about 1~cm worse \wrt the static
case, while the precision is essentially unaltered. As for \crng, the
median error is also the same as in the static case, while the value
of the other metrics is by and large in between the case with all
positions and the one with only \centerpos ones. The precision is
closer to the case of all static positions
(Table~\ref{tab:loc-err-all}), which is mirrored in the slower
increase of the CDF for \crng variants \wrt \sstwr
(Figure~\ref{fig:crng-optitrack-cdf}). Overall, the absolute error is
relatively small and closer to the case with \centerpos positions,
with $\nth{95}\leq 22$~cm. On the other hand, the \nth{99} percentile
is slightly higher than in Table~\ref{tab:loc-err-all}, 
possibly due to the different environment and the 
higher impact of the orientation of the antenna relative to the 
responders.
Another difference \wrt the static case is the
slightly higher precision and \nth{99} accuracy of \ssub \vs
threshold-based estimation, in contrast with the opposite trend we observed
in~\ref{sec:crng-exp-static}. Again, this is likely to be
ascribed to the different environment and MPC profile. 
In any case, this bears only
a minor impact on the aggregate performance, as shown in
Figure~\ref{fig:crng-optitrack-cdf}.

\fakeparagraph{Success Rate} Across the 4,015 signals from our
trajectories, \crng obtained 3,999 position estimates ($99.6\%$) with
both \toa techniques. Nevertheless, 43 of these are affected by an
error $\geq 10$~m and can be disregarded as outliers, yielding an
effective success rate of $98.8\%$, which nonetheless reasserts the
ability of \crng to provide reliable and robust localization.

Regarding ranging, threshold-based estimation yields a success rate of $93.18\%$
across the 24,090 expected estimates, while \ssub reaches $95.4\%$, 
confirming its higher reliability. As expected, the localization
success rate is higher as the position can be computed even 
if several $\hat d_i$ are lost.

\begin{figure}[!tb]
\centering
\includegraphics{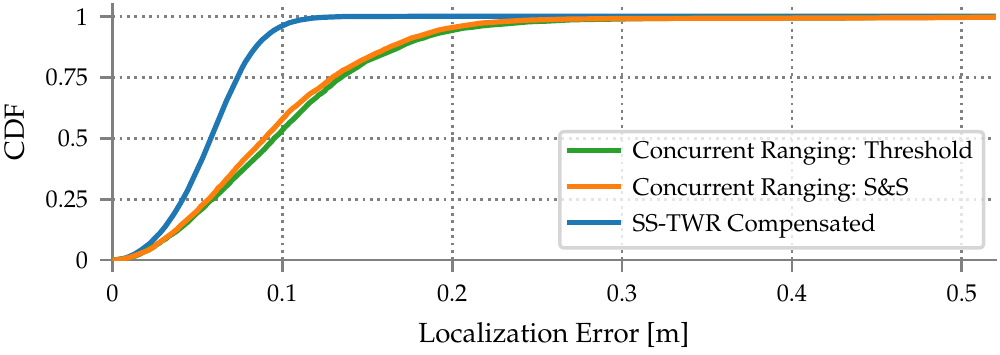}   
\caption{Localization error CDF of \crng \vs compensated \sstwr 
  across multiple trajectories.}
\label{fig:crng-optitrack-cdf}
\end{figure}

\begin{table}
\centering
  \caption{Localization error comparison across multiple mobile trajectories.}
  \label{tab:optitrack-loc-err}
\begin{tabular}{l cc ccccc}
\toprule
& \multicolumn{7}{c}{$\norm{\mathbf{\hat p} - \mathbf{p_r}}$ [cm]}\\
\cmidrule(lr){2-8}
{\bfseries Scheme} & {$\mu$} & {$\sigma$}
  & \nth{50} & \nth{75} & \nth{90} & \nth{95} & \nth{99}\\
\midrule
Concurrent Ranging: Threshold & 12.1 & 17.2 &10 &14 & 18 &22 & 85\\
Concurrent Ranging: \ssub & 11 & 12.8 &9 &13 & 18 &20 & 60\\
\sstwr Compensated & 5.8 & 2.3 &6 &7 & 9 &10 & 12\\
\bottomrule
\end{tabular}
\end{table}

\section{Discussion}
\label{sec:discussion}

The outcomes of our evaluation (\ref{sec:crng-tosn:eval}) across
several static positions and mobile trajectories in two indoor
environments prove that \emph{concurrent ranging reliably provides
  distance and position estimates with decimeter-level accuracy and
  high precision}. The results we presented confirm that concurrent
ranging achieves a performance akin to conventional schemes, and that
it satisfies the strict requirements of most applications, notably
including robot localization.

Nevertheless, \emph{concurrent ranging incurs only a small fraction of
  the cost borne by conventional schemes}. \sstwr requires $2\times N$
packets to measure the distance to $N$ nodes; concurrent ranging
achieves the same goal with a \emph{single} two-way exchange. At the
initiator, often a mobile, energy-bound node, only 2~packets need to
be TX/RX instead of $2\times N$, proportionally reducing energy
consumption and, dually, increasing lifetime. Overall, the ability to
perform ranging via shorter exchanges dramatically reduces channel
utilization and latency, therefore increasing scalability and
update rate. To concretely grasp these claims, consider that, with the
(conservative) response delay $\dtx = \SI{800}{\micro\second}$ we
used, concurrent ranging could provide a location update rate of
$\geq$1,000~Hz, either to a single initiator or shared among
several ones.

Actually achieving these update rates, however, requires a better
hardware and software support than in our prototype. Currently we log
the CIR via USB/UART, as it is the only option with the off-the-shelf
\decawave EVB1000 boards we use. This choice simplifies our prototyping
and enables replication of our results by others, using the same
popular and easily available platform. However, it introduces
significant delays, reducing the location update rate down to only
$\approx$8~Hz; this is appropriate for many applications but
insufficient in others requiring the tracking of fast-moving targets,
\eg drones. Nevertheless, this limitation is easily overcome by
production systems exploiting more powerful and/or dedicated
components, as in the case of smartphones.

Further, this is an issue only if the high update rate theoretically
available must be exploited by a single initiator. Otherwise, when
shared across several ones, our non-optimized prototype could provide its
8~samples per second to $\approx$125~nodes. This would require a
proper scheduling across initiators to avoid collisions, e.g., as
in~\cite{surepoint,talla-ipin}, and incur overhead, ultimately
reducing the final update rate of the system. On the other hand, the
potential for collisions is significantly reduced with our technique,
given that a single concurrent ranging exchange retrieves the
information accrued via $N$ conventional ones. Further, communicating
the schedule could itself exploit concurrent
transmissions~\cite{surepoint,glossy-uwb,uwb-ctx-fire}, opening the
intriguing possibility of merging scheduling and ranging into a single
concurrent exchange abating at once the overhead of both procedures.

Similar issues arise in more dynamic scenarios where ranging is
performed against mobile nodes instead of fixed anchors, e.g., to
estimate distance between humans as in proxemics
applications~\cite{hidden-dimension, proxemic-interactions}. In cases
where the set of nodes is not known a priori, scheduling must be
complemented by continuous neighbor discovery, to determine the set
of potential ranging targets. The problem of jointly discovering,
scheduling, and ranging against nodes has received very little
attention by the research community, although it is likely to become
important for many applications once UWB becomes readily available on
smartphones. In this context, the ability to perform fast and
energy-efficient concurrent ranging against several nodes at once
brings a unique asset, which may be further enhanced by additional
techniques like the adaptive response delays we hinted at
in~\ref{sec:crng-tosn:resp-id}. The exploration of these and other
research avenues enabled by the concurrent ranging techniques we
presented in this paper is the subject of our ongoing work.

Finally, the research findings and system prototypes we describe
in this paper are derived for the DW1000, \ie the only UWB
transceiver available off-the-shelf today. Nevertheless, new
alternatives are surfacing on the market. We argue that the
fundamental concept of concurrent ranging and the associated
techniques outlined here are of general validity, and therefore in
principle transferable to these new transceivers. Moreover, it is
our hope that the remarkable benefits we have shown may inspire new
UWB architectures that natively support concurrent ranging directly
in hardware. 

\section{Related Work}
\label{sec:relwork}

We place concurrent ranging in the context of other UWB ranging
schemes (\ref{sec:relwork:twr}), the literature on concurrent
transmissions in low-power wireless communications
(\ref{sec:relwork:glossy}), and techniques
that build upon the work~\cite{crng} in which we introduced the notion
of concurrent ranging for the first time  (\ref{sec:relwork:crng}).


\subsection{Other UWB Ranging Schemes}
\label{sec:relwork:twr}

Although \sstwr is a simple and popular scheme for UWB, several others
exist, focusing on improving different aspects of its operation.

A key issue is the linear relation between the ranging error and the
clock drift (\ref{sec:toa}). Some approaches extend \sstwr by
\emph{adding} an extra packet from the initiator to the
responder~\cite{polypoint} or from the responder to the
initiator~\cite{ethz-sstwr-drift}. The additional packet 
enables clock drift compensation.

Instead, double-sided two-way ranging (\dstwr), also part of the
\ieeestd standard~\cite{std154}, includes a third packet from the
initiator to the responder in reply to its \response, yielding a more
accurate distance estimate at the responder; a fourth, optional
packet back to the initiator relays the estimate to it.  In the
classic \emph{symmetric} scheme~\cite{dstwr}, the response delay \dtx
for the \response is the same for the third packet from initiator to
responder. 
This constraint reduces flexibility and increases development
complexity~\cite[p. 225]{dw1000-manual-v218}.  In the alternative
\emph{asymmetric} scheme proposed by \decawave~\cite{dw-dstwr-patent,
  dw-dstwr}, instead, the error does not depend on the delays of the
two packets; further, the clock drift is reduced to picoseconds,
making ToA estimation the main source of
error~\cite{dw1000-manual-v218}.  However, \dstwr has significantly
higher latency and energy consumption, requiring up to $4\times N$
packets (twice than \sstwr) to measure the distance to $N$ nodes at
the initiator. We are currently investigating if and how concurrent
ranging can be extended towards \dstwr.

PolyPoint~\cite{polypoint} and SurePoint~\cite{surepoint} improve
ranging and localization by using a custom-designed multi-antenna 
hardware platform. These schemes exploit antenna and channel diversity,
yielding more accurate and reliable estimates; however, 
this comes at the cost of a significantly higher latency 
and energy consumption, decreasing scalability and battery lifetime.

Other schemes have instead targeted directly a reduction of the
packet overhead. The \emph{one-way ranging} in~\cite{ethz-one-way}
exploits \emph{sequential} transmissions from anchors to enable mobile
nodes to passively self-position, by precisely estimating the
time of flight and the clock drift.  However, the update rate and
accuracy decrease as the number $N$ of anchors increases. Other
schemes replace the unicast \poll of \sstwr with a \emph{broadcast}
one, as in concurrent ranging. In N-TWR~\cite{ntwr}, responders send
their \response \emph{sequentially}, to avoid collisions, reducing the
number of packets exchanged to $N + 1$. 
An alternate scheme by \decawave~\cite[p.~227]{dw1000-manual-v218}
exploits a broadcast \poll in asymmetric \dstwr, rather than \sstwr,
reducing the packet overhead to $2 + N$ or $2(N + 1)$ depending on
whether estimates are obtained at the responders or the initiator,
respectively.

In all these schemes, however, the number of packets required grows
linearly with $N$,
limiting scalability. In contrast, concurrent ranging measures the
distance to the $N$ nodes based on a \emph{single} two-way exchange,
reducing dramatically latency, consumption, and channel utilization,
yet providing similar accuracy as demonstrated
in~\ref{sec:crng-tosn:eval}. 


\subsection{Concurrent Transmissions for Low-power Wireless
  Communication}
\label{sec:relwork:glossy}

Our concurrent ranging technique was originally inspired by the body
of work on concurrent transmissions in narrowband low-power radios.
Pioneered by Glossy~\cite{glossy}, this technique exploits the
PHY-level phenomena of constructive interference and capture effect to
achieve unprecedented degrees of high reliability, low latency, and
low energy consumption, as shown by several follow-up
works~\cite{chaos,lwb,crystal}. However, these focus on \ieeestd
narrowband radios, leaving an open question about whether similar
benefits can be harvested for UWB radios.

In~\cite{uwb-ctx-fire} we ascertained empirically the conditions for
exploiting UWB concurrent transmissions for reliable communication,
exploring extensively the radio configuration space. The findings
serve as a foundation for adapting the knowledge and systems in
narrowband towards UWB and reaping similar benefits, as already
exemplified by~\cite{glossy-uwb}. Further, the work
in~\cite{uwb-ctx-fire} also examined the effect of concurrent
transmissions on ranging---a peculiarity of UWB not present in
narrowband---confirming our original findings in~\cite{crng}
(and~\ref{sec:questions}) and analyzing the radio configuration and
environmental conditions in more depth and breadth than what we can
report here.


\subsection{Concurrent Transmissions for Ranging and Localization}
\label{sec:relwork:crng}

We introduced the novel concept of concurrent ranging in~\cite{crng}, 
where we demonstrated the feasibility of exploiting UWB 
concurrent transmissions together with CIR information for ranging; 
\ref{sec:questions} contains an adapted account of the observations
we originally derived. 
Our work was followed by~\cite{crng-graz}, which introduces the idea
of using pulse shapes and response position modulation to match CIR
paths with responders. We discarded the former
in~\ref{sec:crng-tosn:resp-id} and~\cite{chorus} as we verified
empirically that closely-spaced MPC can create ambiguity, and
therefore mis-identifications. Here, we resort to the latter as
in~\cite{chorus, snaploc}, \ie by adding a small time shift $\delta_i$
to each \response, enough to separate the signals of each responder
throughout the CIR span.  The work in~\cite{crng-graz} also suggested
a simpler version of Search \& Subtract for \toa estimation. Instead,
here we follow the original algorithm~\cite{dardari-toa-estimation}
but enforce that candidate paths reach a minimum peak amplitude, to
improve resilience to noise and MPC.  Moreover, we introduce an
alternate threshold-based \toa algorithm that is significantly simpler
but yields similar results. Both preliminary works in~\cite{crng,
  crng-graz} leave as open challenges the TX scheduling uncertainty 
and the unreliability caused by packet loss. Here, we address these
challenges with the local compensation mechanism
in~\ref{sec:crng-tosn:txfix} and the other techniques
in~\ref{sec:reloaded}, making concurrent ranging not only accurate,
but also very reliable and, ultimately, usable in practice.

\decawave~\cite{dw:simulranging} filed a patent on ``simultaneous ranging'' 
roughly at the same time of our original work~\cite{crng},
similarly exploiting concurrent transmissions from responders. 
The patent includes two variants:
\begin{inparaenum}[\itshape i)]
\item a \emph{parallel} version, where all responders transmit nearly simultaneously
  as in~\ref{sec:crng}--\ref{sec:questions}, only aiming to measure
  the distance to the closest responder, and
\item a \emph{staggered} version that exploits time shifts as 
  in~\ref{sec:crng-tosn:resp-id} to determine the distance to each
  responder.
\end{inparaenum}
The latter, however, requires PHY-layer changes that will unavoidably
take time to be standardized and adopted by future UWB transceivers. 
In contrast, the techniques we present here can be exploited with
current transceivers and can also serve as a reference for the 
design and development of forthcoming UWB radios natively 
supporting concurrent ranging. 

Our original paper inspired follow-up work on concurrent
ranging~\cite{crng-graz,R3} but also on other techniques exploiting
concurrent transmissions for localization.  Our own
Chorus~\cite{chorus} system and SnapLoc~\cite{snaploc} 
realize a passive self-localization scheme supporting unlimited
targets. Both systems assume a known anchor infrastructure in which a
reference anchor transmits a first packet to which the others reply
concurrently.  Mobile nodes in range listen for these concurrent
responses and estimate their own position based on time-difference 
of arrival (\tdoa) multilateration. In~\cite{chorus}, we modeled the accuracy of
estimation via concurrent transmissions if the TX uncertainty were to
be reduced, as expected in forthcoming UWB transceivers. This model is
applicable to \crng and, in fact, predicts the results we achieved
in~\ref{sec:crng-tosn:eval} by locally compensating for the TX
uncertainty (\ref{sec:crng-tosn:txfix}). SnapLoc instead proposed to
directly address the TX uncertainty with a correction that 
requires either a wired backbone infrastructure that anchors exploit
to report their known TX error, or a reference anchor that receives
the \response and measures each TX error from the CIR.
Both require an additional step to report the error to 
mobile nodes, and introduce complexity in the deployment along with
communication overhead. 
In contrast, the compensation
in~\ref{sec:crng-tosn:txfix} is \emph{entirely local} to the
responders, therefore imposing neither deployment constraints nor 
overhead. Moreover, the compensation
in~\ref{sec:crng-tosn:txfix} can be directly incorporated
in Chorus and SnapLoc, improving their performance while simplifying
their designs.

Recently, these works have also inspired the use of UWB concurrent
transmissions with angle-of-arrival (AoA)
localization. In~\cite{crng-aoa}, a multi-antenna anchor sends a \poll
to which mobile nodes in range reply concurrently, allowing the anchor
not only to measure their distance but also the AoA of their signals;
combining the two enables the anchor to estimate the position of each
node. The techniques we proposed in this paper (\ref{sec:reloaded})
addressing the TX uncertainty, clock drift, and unreliability caused
by packet loss, are applicable and likely beneficial also for this AoA
technique.

\section{Conclusions}
\label{sec:crng-tosn:conclusions}

In~\cite{crng}, we described the novel concept of concurrent ranging
for the first time in the literature, demonstrated its feasibility,
elicited the open challenges, and outlined the several benefits it
could potentially enable in terms of latency, scalability, update
rate, and energy consumption.

In this paper, we make these benefits a tangible reality. We tackle
the aforementioned challenges with a repertoire of techniques that,
without requiring modifications to off-the-shelf UWB transceivers,
turn concurrent ranging into a practical and immediately available
approach. Concurrent ranging empowers the designers of UWB ranging and
localization systems with a new option whose accuracy is comparable to
conventional techniques, but comes at a fraction of the latency and
energy costs, therefore unlocking application trade-offs hitherto
unavailable for these systems.

\begin{acks} 
  This work is partially supported by the Italian government via the
  NG-UWB project (MIUR PRIN 2017).
  We are grateful to Jarek Niewczas from \decawave for his detailed
  suggestions on how to mitigate the TX scheduling uncertainty.  We
  also wish to acknowledge the help of our collaborators: Davide
  Vecchia for helping us with the implementation, as well as Timofei
  Istomin and Davide Molteni for their support with experiments.
\end{acks}

\bibliographystyle{ACM-Reference-Format}

\end{document}